\RequirePackage{fix-cm}
\documentclass[twocolumn]{svjour3}          
\smartqed  
\usepackage[pdftex]{graphicx}
\usepackage{mathptmx}      
\usepackage[utf8]{inputenc}
\usepackage[T1]{fontenc}

\usepackage{tikz}
\usepackage{tikz-qtree}
\usetikzlibrary{shapes,arrows,trees,positioning}
\usetikzlibrary{decorations.pathmorphing,decorations.pathreplacing}

\tikzset{
		edge from parent/.style= 
            {line width=0.5pt, draw, edge from parent fork right},
         every tree node/.style={anchor=west, inner ysep=2pt, align=center, minimum height=14pt},
         grow'=right,
         level distance=90pt,
         sibling distance=0pt
		}
\usepackage{subfigure}
\usepackage{color}
\usepackage[hyphens]{url}
\usepackage{hyperref}
\hypersetup
{
    pdfauthor={Harald Lampesberger},
    pdftitle={Technologies for Web and cloud service interaction: a survey},
    citecolor=blue,
    urlcolor=black,
    linkcolor=black,
    colorlinks=true
}
\usepackage[final]{listings}
\journalname{Service Oriented Computing and Applications}

\begin{document}

\author{Harald Lampesberger}

\institute{H. Lampesberger \at
              Christian Doppler Laboratory for Client-Centric Cloud Computing,\\
              Johannes Kepler University Linz,\\
              Softwarepark 21, 4232 Hagenberg, Austria\\
              \email{h.lampesberger@cdcc.faw.jku.at\\
              \\
              This is the author's accepted draft. Please cite the published version: Lampesberger, H.: Technologies for Web and cloud service interaction: a survey. Service Oriented Computing and Applications (2015). DOI 10.1007/s11761-015-0174-1}}        

\title{Technologies for Web and cloud service interaction: a survey}
\date{Received: date / Accepted: date}

\maketitle

\begin{abstract}
The evolution of Web and service technologies has led to a wide landscape of standards and protocols for interaction between loosely coupled software components.
Examples range from Web applications, mashups, apps, and mobile devices to enterprise-grade services.
Cloud computing is the industrialization of service provision and delivery, where Web and enterprise services are converging on a technological level.
The article discusses this technological landscape and, in particular, current trends with respect to cloud computing.
The survey focuses on the communication aspect of interaction by reviewing languages, protocols, and architectures that drive today's standards and software implementations applicable in clouds.
Technological advances will affect both client side and service side.
There is a trend toward multiplexing, multihoming, and encryption in upcoming transport mechanisms, especially for architectures, where a client simultaneously sends a large number of requests to some service.
Furthermore, there are emerging client-to-client communication capabilities in Web clients that could establish a foundation for upcoming Web-based messaging architectures.

\keywords{Web technology \and Web services \and Cloud services \and Service architecture \and Communication protocols \and Languages \and Service interaction patterns}
\end{abstract}

\section{Introduction}
\label{sec:Introduction}

The need to share network-based resources and use remote functionality in a program without dealing with low-level network access has fueled discussions in the 1970s \cite{RFC674,RFC684,RFC707} and ultimately led to the remote procedure call (RPC) framework by Birrell and Nelson \cite{Birrell1984} in the 1980s.
RPC became a driver in enterprise systems; location transparency of procedures eases code reuse but requires tight coupling, e.g., a unified type system.
In the 1990s, the principles of object orientation and RPC gave raise to distributed objects \cite{omg-corba}.
Tight coupling and interaction complexity in RPC and distributed objects affected the scalability of enterprise systems, and at the end of the 1990s, message passing between so-called services became an alternative enterprise architecture with relaxed coupling and easier scalability.
Today, middleware for message queuing and the concept of service-oriented architecture (SOA)~\cite{oasis-soa} dominate large-scale distributed enterprise systems.

In the meantime, Berners-Lee \cite{w3c-history} laid out the foundation for a World Wide Web of nonlinear text documents, i.e., hypertext, exchanged in a client-server architecture over the predecessor of today's Internet in 1989.
The first Web browser was announced end of 1990, the World Wide Web Consortium (W3C) was established in 1994, and W3C published the first Hypertext Markup Language (HTML) Recommendation in 1997.
Since then, the Web has evolved from simple hypermedia exchange, to interactive user interfaces, rich client applications, user-provided content, mashups, social platforms, and wide-scale mobile device support.
Web technology has become pervasive and is not limited to hypermedia applications anymore.
Standards are widely accepted, and they have contributed to the success of Web services because protocols like the Hypertext Transfer Protocol (HTTP) are reliably forwarded over the Internet~\cite{Alonso2004,Pautasso2008}.

Cloud computing \cite{Armbrust2010} can be seen as the industrialization of service provision and delivery over the Internet using established technologies from the Web \cite{Foster2008} and from enterprise services \cite{Bernstein2009}.
The goal is to offer a service, accessible across devices, systems, and platforms.
Interaction by communication between service consumers and providers, i.e., clients and services, is therefore a key aspect of service delivery.
Cloud service delivery models like Platform-as-a-Service (PaaS) or Software-as-a-Service (SaaS) benefit from well-known Web standards and their wide acceptance~\cite{Armbrust2010,Foster2008,Schewe2011a}.
To acknowledge this relation between Web and services technology, this article surveys the state-of-the-art and recent trends in technologies applicable in clouds.

\subsection{Scope}

A survey of technologies in such a dynamic environment needs a defined scope.
All technologies that allow a client to interact with a service should be considered; however, the notion of \emph{service} is not precisely defined~\cite{Schewe2011a}.
The following informal properties and restrictions therefore characterize a service in context of this work:

\begin{itemize}
	\item \emph{Service interface.} \ Services are considered as distributed, network-accessible software components that offer functionality and need communication for interaction \cite{Gottschalk2002,Pautasso2008}.	
	A notion of interface that accepts a certain language is therefore required.
	The survey is restricted to technologies that enable communication between clients and service interfaces applicable in Web, PaaS, and SaaS cloud delivery models.
	\item \emph{Heterogeneous platforms.} \ A characteristic of service-orientation is to provide functionality and content across hard- and software platforms.
	Only technologies that embrace this compatibility are considered. 
	\item \emph{Publicly available standards.} \ The focus is on technologies that are available to the public audience, in particular, technologies based on Internet protocols, i.e., the TCP/IP protocol suite \cite{Stevens1993}, and with publicly available specifications.
	Specialized technologies for a limited audience or application, like industrial control systems, are not part of this study.
	\item \emph{Parties.} \ There are two participating parties or peers in service interaction: a \emph{client} that consumes some \emph{service} offered by a provider or server, i.e., client-to-service interaction.
	On a conceptual level, a service can participate also as a client to consume other services for a composition, i.e., service-to-service interaction.
	Furthermore, a service can coordinate two clients to establish client-to-client or peer-to-peer interaction.
\end{itemize}

In accordance with the aforementioned characteristics, the state-of-the-art and recent trends in Web and service communication technologies applicable to the Web, PaaS, and SaaS are surveyed.

\subsection{Motivation}
\label{sec:motivation}

This survey is motivated by ongoing research efforts in formal modeling of cloud services \cite{Bosa2013,Bosa2013b,Schewe2011a}, modeling of service quality \cite{Rady2014,Rady2013}, service adaptation~\cite{Chelemen2013}, identity management \cite{Vleju2014,Vleju2012c}, and security monitoring \cite{Lampesberger2014,Lampesberger2013b,Lampesberger2012} for cloud services.
All these aspects need communication between clients and services.
Understanding the state-of-the-art in service communication is therefore necessary, e.g., for security research because an ambiguous or imprecise service interface is in fact a gateway for attacks~\cite{Sassaman2013}.

There is a rich body of literature using \emph{patterns} to describe service interaction on a conceptual level \cite{Aalst2009,Barros2005a,Barros2005,Hohpe2003,Zaha2006}.
On the other hand, the numerous software implementations used in today's services are heavily driven by continuously evolving standards and ad hoc specifications.
This work aims to bridge this gap by surveying the state-of-the-art of technologies and resort to patterns when concepts are discussed.
Patterns are appealing because they allow to describe solutions in a conceptual way and can therefore support service integrators and scientists in understanding new technologies.

\subsection{Methodology and structure}
\label{sub:methodology}

\begin{figure}
	\begin{center}
\begin{tikzpicture}[->, >=latex]
\node (v1) at (0.5,0) {\textsf{Protocol}};
\node (v3) at (2.6,-1.1) {\textsf{Language}};
\node (v2) at (5,0) {\textsf{Architecture}};
\node (v4) at (5,-1.1) {\textsf{Implementation}};
\node (v5) at (2.6,1) {\textsf{Service Interaction Pattern}};
\draw  (v1) edge (v2);
\draw  (v3) edge (v1);
\draw  (v3) edge (v2);
\draw  (v2) edge (v4);
\draw  (v1) edge[loop left] (v1);
\draw[<->]  (v5) edge (v1);
\draw[<->]  (v5) edge (v2);
\draw[dotted]  (-0.4,0.5) rectangle (1.4,-2);
\draw[dotted]  (3.8,0.5) rectangle (6.2,-2);
\draw[dotted]  (1.5,0.5) rectangle (3.7,-2);
\draw[dotted]  (-0.4,1.4) rectangle (6.2,0.6);
\node at (0.5,-1.7) {Section \ref{sec:protocols}};
\node at (5,-1.7) {Section \ref{sec:architectures}};
\node at (2.6,-1.7) {Section \ref{sec:languages}};
\node at (5.4,1) {Section \ref{sec:patterns}};
\end{tikzpicture}
	\end{center}
\caption{The concepts and their relationships found in communication technologies also reflect in the structure of the survey}
\label{fig:concepts}
\end{figure}
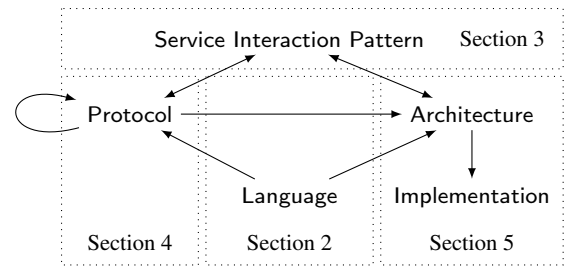

The problem is approached both in top-down and bottom-up manners.
In the top-down view, entities of information are exchanged between peers, i.e., clients and services, in an agreed-upon style, i.e., interaction pattern.
The article therefore discusses languages for encapsulation information, so information becomes transportable.
The bottom-up view investigates how an agreed-upon style of interaction is actually implemented in today's networks, i.e., communication protocols and architectures for information exchange.
Fig. \ref{fig:concepts} visualizes the relationships between concepts required for service interaction. 
The relations and concepts have been derived from the extensive literature review in this article.

\emph{Languages} are fundamental for communication.
A language defines an alphabet, syntax, and semantics to represent information in a transportable format.
Languages therefore encode content, media, and information in general.
Popular languages in the Web and for services are discussed in Sect. \ref{sec:languages}.

Due to the many styles of interaction, conceptual patterns for a unified nomenclature are considered, in particular, \emph{service interaction patterns} (Sect. \ref{sec:patterns}) introduced by Barros et al. \cite{Barros2005,Barros2005a} in context of the Workflow Patterns Initiative \cite{workflowpatterns}.
These patterns are rather informal but manageable in their number and sufficiently abstract to discuss and compare interaction in the survey.
The relations between patterns, protocols, and architectures are bidirectional from a historic point of view; patterns have been derived from successful protocols, and patterns have influenced the specification of new protocols.

A communication \emph{protocol} specifies a language for a communication channel and rules of engagement to implement a certain interaction pattern.
Protocols can integrate other protocols to become a protocol stack, as common in Internet protocols, with the self-reference in Fig. \ref{fig:concepts} representing this relation.
Relevant protocols for Web and cloud services are surveyed in Sect. \ref{sec:protocols}.

To deliver a service, an \emph{architecture} specifies communication protocols for low-level interaction, languages for encoding information, and high-level interaction patterns between peers.
Architectures are basically blueprints for available transport mechanisms and formats, and Sect. \ref{sec:architectures} investigates architectures found in the Web and services applicable to cloud computing.
Architectures are eventually implemented as executable software, and popular implementations are discussed when appropriate.
Scientific findings, observations, and potential implications are then discussed in Sect. \ref{sec:discussion}, whereas Sect. \ref{sec:conclusion} concludes the survey.

The survey investigates text-based, binary, and container formats for information encoding; protocols in terms of the TCP/IP protocol stack, including multiplexing and multihoming transport mechanisms, HTTP extensions, and wire formats for messaging protocols; and architectures in a Web-oriented view (Web applications, Web syndication, and Web mashups) and a service-oriented view (RPC, Web services, and messaging solutions).
The contribution of this article is a discussion of cloud aspects and cloud-specific developments; multiplexing, multihoming, and encryption in modern transport mechanisms; correctness of content types; upcoming client-to-client capabilities; and the impact on network traffic monitoring.

\subsection{Standardization bodies}

Standards for languages, protocols, and architectures in the Internet, in the Web, and for services are primarily driven by nonprofit organizations, communities, consortia but also enterprises.
Important institutions are therefore recalled.

To develop industrial standards on a global scale such as specifications for electronic communication devices, e.g., networking, the International Organization for Standardization (ISO) \cite{iso}, the International Electrotechnical Commission (IEC) \cite{iec}, and the International Telecommunication Union (ITU) \cite{itu} are three connected organizations that closely work together.
The Institute of Electrical and Electronics Engineers (IEEE) Standards Association \cite{ieee} is also well known for global networking standards, e.g., the IEEE 802.3 Ethernet standard.

With respect to language and protocol specifications, the Internet Engineering Task Force (IETF) \cite{ietf} organizes a community process to develop standards, especially communication protocols, through open Request for Comments (RFCs).
The W3C \cite{w3c} drives Web standardization efforts, and the Object Management Group (OMG) \cite{omg} aims for standardized business process modeling.
Another nonprofit organization for developing open standards for languages and protocols, specifically for enterprise services, is the Organization for the Advancement of Structured Information Standards (OASIS)~\cite{oasis}.

The Internet Corporation for Assigned Names and Numbers (ICANN) \cite{icann}, including its department Internet Assigned Numbers Authority (IANA) \cite{iana}, is a nonprofit organization for directing worldwide the agreement on Internet addresses, domain names, and protocol identifiers.
Also, a number of standards have been proclaimed by enterprises that use them internally or offer them as software or services, e.g., Amazon, Cisco Systems, Google, Facebook, IBM, Microsoft, and Oracle.

\section{Languages for content and media}
\label{sec:languages}

Formally, a language is a (possibly infinite) set of strings generated from a finite set of symbols, referred to as alphabet \cite{Hopcroft2000}.
Languages are essential to communicate information represented as messages.
While information exchange in Web and cloud services can be distinguished into message and stream based, a stream is in fact a single message sent in chunks or as a sequence of individual smaller messages.
Languages for encoding content or media are also referred to as data serialization formats or formats in short.

Communicating parties can only parse content of a certain kind, where the format, i.e., syntax, and meaning, i.e., semantics, of the language are defined.
The hardness of parsing is then a computational complexity property of the language \cite{Grune2010}: With increased expressiveness, more and more information can be encoded in a language, but parsing also becomes harder and therefore more error-prone in software implementations \cite{Sassaman2013}.

Alphabets for intercommunicating digital systems are typically binary, and the basic unit of information is a \emph{bit}.
For Internet applications, a \emph{byte} of eight bits is a common transferable unit.
Content can be distinguished into binary and text based with respect to the alphabet:

\begin{itemize}
	\item \emph{Binary content.} \ When a language describes a bijection between digital sequences and the domain of actual values and structures, then contents are referred to as binary content and they are likely not human-readable.
	\item \emph{Text-based content.} \ 	Text is not simply text, but rather bits and bytes with an associated mapping to human-readable symbols, so some digital sequence has a textual representation.
	Such a mapping is called \emph{character encoding} or character set, e.g., ASCII.
	Content is said to be text based, if its syntax has a human-readable representation.
\end{itemize}

ASCII is the most fundamental character encoding; it uses seven bits to enumerate a set of control and printable characters, but it is limited to the English alphabet.
Unicode~\cite{unicode} attempts to enumerate all the human-readable symbols in all natural languages.
Character encodings like the ASCII-compatible UTF-8~\cite{RFC3629} then specify a compact, byte-oriented encoding to represent millions of symbols efficiently.

\subsection{Content types}

Formally, a type is a general concept shared by a set of objects, also referred to as the instances of a type.
With respect to content and media, a content type is then an identifier that specifies the alphabet, character encoding, syntax, and eventually semantics of a language.
The notion of content type is essential for modular software design, where an appropriate parser is chosen during runtime based on the content type.

In today's applications, the Multipurpose Internet Mail Extensions (MIME) have become an Internet standard for specifying content types in the Web \cite{RFC2046}.
They are referred to as \emph{MIME content type}, \emph{Internet media type}, or \emph{MIME type} for short.
For example, the MIME type of a simple ASCII text is \texttt{text/plain}.
A MIME type identifier can also explicitly refer to the character encoding of text-based contents \cite{RFC6657}, e.g., \texttt{text/plain; charset=utf-8}.

\subsection{Text-based content}
\label{sub:textcontent}

Compared to binary formats, text-based languages have a lower information density because human-readable symbols need to be digitally encoded.
For example, an integer number $n > 0$ needs $\lceil \log_{10} n \rceil$ bytes in UTF-8 encoding, while efficient binary representation requires only $\lceil \log_{2} n\rceil$ bits.
Despite lower information density, the worldwide agreement on human-readable encodings has contributed the success of hypermedia.

\subsubsection{Semi-structured languages}

Three of the most influential languages for information exchange in the Web are HTML, the Extensible Markup Language (XML), and the JavaScript Object Notation (JSON).

\paragraph{Hypertext Markup Language.} \ HTML is the standard for defining websites and has the MIME type \texttt{text/html}.
It uses markup such as tags, attributes, declarations, or processing instructions to express structural, presentational, and semantic information as text.
While earlier HTML versions up to 4.01 \cite{w3c-html4} are an application of the Standard Generalized Markup Language (SGML), which requires a complex SGML parser framework, today's HTML5 \cite{w3c-html5} specifies an individual parser.

\begin{figure}
\centering
\begin{lstlisting}
<!DOCTYPE html>
<html lang="en">
  <head>
    <meta charset="utf-8">
    <title>Hello World</title>
  </head>
  <body>
    <h1>Headline</h1>
    <p>A sample paragraph.</p>
    <script>alert('Hello, World!')</script>
  </body>
</html>
\end{lstlisting}
\caption{A minimal example for HTML5 has the characteristic document type declaration and a \texttt{html} root tag}
\label{fig:html}
\end{figure}

\ \ An examplary HTML5 document is listed in Fig. \ref{fig:html}.
\mbox{SGML-based} parsers distinguish the grammars of different HTML versions in the document type declaration in the first line.
As HTML5 is not SGML based, the document type declaration is deliberately incomplete to indicate SGML independence.
A document is separated into a \texttt{header} for metadata and a \texttt{body} for the semi-structured content of a website.
All allowed tags are specified in the standard.
Interestingly, the character encoding of a document is defined within the document itself in a \texttt{meta} tag.
This tag should be the fist tag in the header, so the parser becomes aware of the encoding before other tags are encountered.

An SGML or HTML5 parser in a Web browser transforms a document into a Document Object Model (DOM) \cite{w3c-dom}, a generalized tree-like data structure that is eventually rendered visible by the user interface.
Another popular format with respect to HTML is Cascading Style Sheets (CSS) \cite{w3c-css} for defining both the look and behavior of a DOM's visual representation.

\paragraph{Extensible Markup Language.} \ XML \cite{w3c-xml} originates from SGML, but it is more restricted and popular for electronic data exchange.
An example is shown in Fig. \ref{fig:xml}.
Tags, attributes, namespaces, declarations, and processing instructions are syntactic constructs for structuring information in XML as text.
The first line in an XML document should be a processing instruction that informs the parser about the XML version and the applied character encoding.

XML is a language family: The structuring of elements (tag names) and attributes within a document is unrestricted, and only the syntactic rules have to be obliged.
Element content is limited to text; by default, XML distinguishes two datatypes, parsed (PCDATA) and unparsed character data (CDATA).
\emph{Mixed-content} XML relaxes element content restrictions; text in element content is also allowed to nest other elements, e.g., the \texttt{review} element in Fig. \ref{fig:xml}.

\begin{figure}
\centering
\begin{lstlisting}
<?xml version="1.0" encoding="UTF-8"?>
<movie year="1968">
  <title>2001: A Space Odyssey</title>
  <director nid="nm0040">S. Kubrick</director>
  <review>A <em>good</em> movie.</review>
</movie>
\end{lstlisting}
\caption{An exemplary XML document with mixed-content \cite{Lampesberger2013b}}
\label{fig:xml}
\end{figure}

The underlying logical structure of XML is a tree; therefore, open and close tags must be correctly nested.
A document with correct nesting, proper syntax, and a single root element is \emph{well-formed}.
Furthermore, a document is said to have an XML Information Set \cite{w3c-infoset} if it is well-formed and namespace constraints are satisfied.
An Infoset is an unambiguous abstraction from textual syntax; e.g., there are two syntactic notions for empty elements in plain XML.

To restrict the structure, XML offers schema languages, e.g., Document Type Definition (DTD)~\cite{w3c-xml}, XML Schema (XSD)~\cite{w3c-xmlschema}, and Relax NG~\cite{oasis-relaxng}.
Formally, a schema is a grammar that characterizes a set of XML documents, and a document is said to be \emph{valid} if its schema is obeyed \cite{Martens2006}.
In this sense, a schema allows to specify a content subtype of XML.
XSD and Relax NG also support more fine-grained datatypes than PCDATA and CDATA for restricting element contents.

The duality of text representation and logical tree structure of documents has led to two different processing approaches.
A document can be either parsed into a DOM for tree operations or processed directly as a stream of text, open-, or close-tag events using the Simple API for XML (SAX) \cite{saxproject}.

Repeated element names in tags reduce the information density in XML.
SXML \cite{Kiselyov2002} is an alternative syntax using S-expressions such that element names occur only once, and higher information density is achieved.

The MIME media type of the XML language family is \texttt{application/xml}.
XHTML \cite{w3c-xhtml} is a re-specification of HTML and has a MIME type that indicates the XML origin (\texttt{application/xhtml+xml}).
XHTML is conceptually an XML subtype with a strict syntax specified in a schema, so an XML parser can be used instead of a complex markup parser.
XML is also the supertype for many Web formats, e.g., Scalable Vector Graphics (SVG) \cite{w3c-svg} or Mathematical Markup Language (MathML) \cite{w3c-mathml}, and MIME types for well-known subtypes are specified in RFC 3023 \cite{RFC3023}.

\paragraph{JavaScript Object Notation.} \ JSON \cite{RFC7159} is a simple text format to serialize information as structured key-value pairs.
JSON has the MIME type \texttt{application/json}, and it is human-readable, as shown in Fig. \ref{fig:json}.
The syntax is a subset of the JavaScript language (discussed in Sect. \ref{sub:JavaScript}), and a JSON document is either parsed or evaluated to an object during runtime.
JSON specifies six basic datatypes: \texttt{null}, \texttt{Number}, \texttt{String}, \texttt{Boolean}, \texttt{Array}, and \texttt{Object}, and syntactic rules to represent them as text. 
A proper JSON document always has a single root object.

\begin{figure}
\centering
\begin{lstlisting}
{ "movie": {
  "year": 1968,
  "title": "2001: A Space Odyssey",
  "director": {"nid": "nm0040", "n": "S. Kubrick"},
  "review": [
    {"text": "A good movie", "public": true},
    {"text": "Complex", "public": false}
  ],
  "prequel": null } }
\end{lstlisting}
\caption{A JSON document that applies all available basic datatypes and has a length of 248 bytes in UTF-8 encoding}
\label{fig:json}
\end{figure}

Similar to XML, JSON defines a family of languages because there are syntactical restrictions, but no structural limitations in the standard.
JSON Schema \cite{ietf-jsonschema} is a schema language expressed in JSON format, and the motivation is the same as in XML schemas: to define a set of JSON documents and enable schema validation.

\subsubsection{Binary-to-text encodings}

Base16, Base32, Base64 \cite{RFC4648}, Per\-cent-Encoding \cite{RFC3986}, and Quoted-Printable \cite{RFC2045} are binary-to-text encoding schemes to map an arbitrary binary value to a text of ASCII printable characters.
Most notably, Base64 encoding is a popular method to embed arbitrary binary content in XML or JSON as text, but incurs a $33\%$ overhead due to reduced information density.

\subsubsection{Other text-based formats}

There exist several specifications for text-based languages with varying popularity, for example, comma-\-se\-pa\-rated values (CSV) \cite{RFC4180} for relational data; markup languages Candle \cite{candle} and YAML \cite{yaml}; the Ordered Graph Data Language (OGDL) \cite{Veen2014} for graph-structured data; and the Open Data Description Language (OpenDDL) \cite{Lengyel2013} for structured information.

\subsection{Binary content}
\label{sub:binary content}

Binary formats offer compact representation of information but are are typically not human-readable, e.g., audio and video formats.
Abstract Syntax Notation One (ASN.1)~\cite{Dubuisson2001} is a popular standard for structured information exchange.
ASN.1 distinguishes between an abstract specification of information structure and encoding rules such as Basic Encoding Rules (BER) or Distinguished Encoding Rules (DER).
An encoding rule defines how an actual instance, according to some abstract structure, translates to bits and bytes.
An application of ASN.1 using DER are X.509 certificates in today's public key infrastructures \cite{Rescola2001}.

As text-based languages incur overhead, binary equivalents to popular text-based formats have been proposed.
With respect to XML, binary representations are: Efficient XML Interchange (EXI) \cite{w3c-exi}; .NET Binary XML \cite{ms-binaryxml}; Fast Infoset \cite{itu-fi} as an application of ASN.1; and Binary MPEG Format for XML (BiM) \cite{iso-bim}, which originates from a video format.

There are also attempts to optimize JSON representation through binary equivalents, e.g., Binary JSON \cite{bson}, MessagePack \cite{msgpack}, and Concise Binary Object Representation (CBOR) \cite{RFC7049}.
Fig. \ref{fig:messagepack} gives an example, how MessagePack translates the JSON document from Fig. \ref{fig:json} into a more compact form.

\begin{figure}
\centering
\begin{lstlisting}
81 a5 6d 6f 76 69 65 85 a4 79 65 61 72 cd 07 b0 a5 74
69 74 6c 65 b5 32 30 30 31 3a 20 41 20 53 70 61 63 65
20 4f 64 79 73 73 65 79 a8 64 69 72 65 63 74 6f 72 82
a3 6e 69 64 a6 6e 6d 30 30 34 30 a1 6e aa 53 2e 20 4b
75 62 72 69 63 6b a6 72 65 76 69 65 77 92 82 a4 74 65
78 74 ac 41 20 67 6f 6f 64 20 6d 6f 76 69 65 a6 70 75
62 6c 69 63 c3 82 a4 74 65 78 74 a7 43 6f 6d 70 6c 65
78 a6 70 75 62 6c 69 63 c2 a7 70 72 65 71 75 65 6c c0
\end{lstlisting}
\caption{A MessagePack equivalent of the JSON document in Fig. \ref{fig:json}. The binary representation in hexadecimal notation reduces the length to 144 bytes \cite{msgpack}}
\label{fig:messagepack}
\end{figure}

Some RPC architectures (Sect. \ref{sub:rpc}) specify binary formats to serialize data, e.g., External Data Representation (XDR)~\cite{RFC1014}, Apache Avro~\cite{apache-avro}, Apache Etch~\cite{apache-etch}, Apache Thrift~\cite{apache-thrift}, Protocol Buffers~\cite{google-protobuf}, and Hessian~\cite{caucho-hessian}.
Other binary languages worth mentioning are the Structured Data Exchange Format (SDXF) \cite{RFC3072} for hierarchical structures, and Property List \cite{apple-propertylist}, a data serialization format in Apple systems.

\subsection{Container formats}
\label{sub:containers}

A container is an encoding to encapsulate other arbitrary contents.
The MIME standards specify a \emph{multipart} \cite{RFC2046} content type for containers, where contents of varying types are interleaved.
In multipart, a text-based boundary string separates individual parts as shown in Fig. \ref{fig:multipart}. 
Each part has a text-based header that denotes its \texttt{Content-Type} and additional metadata such as binary-to-text or transfer encodings.
Multipart defines several subtypes:

\begin{itemize}
	\item \texttt{multipart/alternative} \cite{RFC2046} to model a choice over multiple contents to a consumer;
	\item \texttt{multipart/byteranges} \cite{RFC2616} to encapsulate a subsequence of bytes that belongs to a larger message or file;
	\item \texttt{multipart/digest} \cite{RFC2046} to store a sequence of text-based messages;
	\item \texttt{multipart/form-data} \cite{RFC2388} for submitting a set of completed form fields from an HTML website;
	\item \texttt{multipart/message} \cite{RFC2046} for an email message;
	\item \texttt{multipart/mixed} for inline placement of media in a text-based message, e.g., embedded images in emails;
	\item \texttt{multipart/parallel} \cite{RFC2046} to process all parts simultaneously on hardware or software capable of doing so;
	\item \texttt{multipart/related} \cite{RFC2387} as a mechanism to aggregate related objects in a single content;
	\item \texttt{multipart/report} \cite{RFC6522} as a container for email messages; and
	\item \texttt{multipart/x-mixed-replace} \cite{netscape-mime} for a stream of parts, where the most recent part always invalidates all preceding ones, e.g., an event stream.
\end{itemize}

\begin{figure}
\centering
\begin{lstlisting}
Content-Type: multipart/mixed; boundary="endpart"

Data in this section is undefined.
--endpart
Content-Type: text/plain; charset=utf-8

This first part is UTF-8 encoded text.
--endpart
Content-Type: image/gif
Content-Transfer-Encoding: base64

R0lGODlhAQABAIAAAAUEBAAAACwAAAAAAQABAAACAkQBADs=
--endpart--
\end{lstlisting}
\caption{This text-based MIME \texttt{multipart/mixed} container example has two parts and demonstrates the boundary principle}
\label{fig:multipart}
\end{figure}

An application of multipart is XML-binary Optimized Packaging (XOP) \cite{w3c-xop}.
XOP specifies a container format for XML as a MIME \texttt{multipart/related} package \cite{RFC2387}, where binary element content is directly embedded to remove the necessity of binary-to-text encoding.
The W3C specifies a set of attributes and rules for XML and XSD to handle MIME types \cite{w3c-dmcbdx} which are effectively used in XOP. 

S/MIME \cite{RFC5751} is a security extension for MIME.
It defines encryption (\texttt{multipart/encrypted}) and digital signatures (\texttt{multipart/signed}) for confidentiality, integrity, and non-repudiation of data using public key cryptography.

In general, a drawback of MIME multipart is that the boundary string must not appear in any of the parts because it would break the format.
Microsoft's Direct Internet Message Encapsulation (DIME) \cite{ms-dime} is another standard for encapsulation and streaming of arbitrary binary data in the spirit of MIME multipart, but with an improved boundary mechanism to reliably distinguish parts.

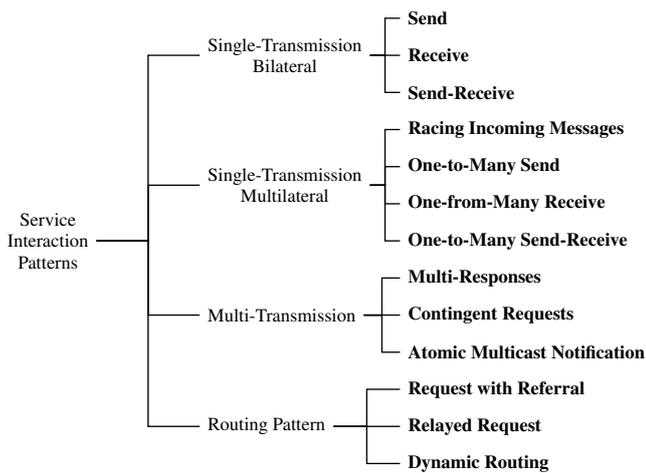
\begin{figure}
	\begin{center}
\begin{tikzpicture}[level distance=75pt, font=\scriptsize]
\tikzset{level 1/.style={level distance=65pt}}
\Tree[.{Service\\Interaction\\Patterns}
	[.{Single-Transmission\\Bilateral}
		{\textbf{Send}}
		{\textbf{Receive}}
		{\textbf{Send-Receive}}
	]
	[.{Single-Transmission\\Multilateral}
		{\textbf{Racing Incoming Messages}}
		{\textbf{One-to-Many Send}}
		{\textbf{One-from-Many Receive}}
		{\textbf{One-to-Many Send-Receive}}
	]
	[.{Multi-Transmission}
		{\textbf{Multi-Responses}}
		{\textbf{Contingent Requests}}
		{\textbf{Atomic Multicast Notification}}
	]
	[.{Routing Pattern}
		{\textbf{Request with Referral}}
		{\textbf{Relayed Request}}
		{\textbf{Dynamic Routing}}
	]
]
\end{tikzpicture}
	\end{center}
\caption{The service interaction patterns by Barros et al. \cite{Barros2005,Barros2005a} characterize message-based interaction between two or more parties}
\label{fig:interaction_patterns}
\end{figure}

\subsection{JavaScript}
\label{sub:JavaScript}

ECMAScript \cite{ecmascript}, better known as JavaScript, is a dynamic programming language which is supported by practically all available Web browsers today.
JavaScript can be regarded as text-based syntax and a JavaScript runtime environment.
A runtime environment is typically present in a Web browser but also available for services, e.g., Node.js \cite{nodejs}.

JavaScript code is either embedded directly into HTML or XHTML markup using \texttt{script} tags, inlined in attributes, or exchanged as text-based content.
When exchanged as resource, JavaScript code has an individual MIME type, i.e., \texttt{application/javascript} \cite{RFC4329}.
A browser runtime environment then interprets the code to dynamically adapt the DOM and its visual representation.
While JavaScript is in fact a Turing-complete programming language, its runtime environment enforces restrictions on system and network access during execution, e.g., local file access, to enforce a security policy.

\section{Service interaction patterns}
\label{sec:patterns}

Clients and services exchange messages of certain content types in an interaction.
The service interaction patterns introduced by Barros, Dumas, and Ter Hofstede~\cite{Barros2005,Barros2005a} are recalled in Fig. \ref{fig:interaction_patterns}.
They characterize styles of message exchange between parties in a distributed network and  propose three dimensions to distinguish patterns:
the number of participating parties; the number of message transmissions in an interaction; and whether messages in two-way interaction are routed to third-parties or take round trips.
This leads to four pattern groups: single-transmission bilateral, single-transmission multilater\-al, multi-transmission, and routing patterns.
This section is a high-level summary of the original descriptions \cite{Barros2005a}.

\subsection{Single-transmission bilateral patterns}

\begin{description}
	\item[{Send}.] Party $A$ sends party $B$ a message.
	
	Related to: unicast, point-to-point send.

	\item[{Receive}.] Party $A$ awaits an incoming message.
	
	Related to: listener, event handler.
	
	\item[{Send-Receive}.] Party $A$ sends party $B$ a message and causes $B$ to respond with a message.
	This pattern has a dual, i.e., Receive-Send, where party $A$ waits for an incoming message and returns a response when received.
	
	Related to: request-response, request-reply, RPC.
\end{description}

\subsection{Single-transmission multilateral patterns}
\begin{description}
	\item[{Racing Incoming Messages}.] Party $A$ waits for a single incoming message from a number of possible messages and senders.
	The continuation of $A$ after receiving the first message depends on the message type or sender, and subsequent messages may or may not be discarded.
	
	Related to: racing messages, deferred choice.
	
	\item[{One-to-Many Send}.] Party $A$ sends $n$ messages to the parties $B_1 \dots B_n$, where all messages have the same type but not necessarily the same content.
	
	Related to: multicast, broadcast (where a virtual broker addresses all parties in the domain), event notification, scatter, publish-subscribe, fan-out.

	\item[{One-from-Many Receive}.] Party $A$ awaits a number of logically connected messages from senders $B_1 \dots B_n$. The messages have to arrive within a certain time span so they can be linked together.
	
	Related to: event aggregation, gather, fan-in.

	\item[{One-to-Many Send-Receive}.] Party $A$ sends $n$ request messages to recipients $B_1 \dots B_n$ and waits for a certain time span for responses from the recipients.
	Whether $A$ considers the interaction successful or not depends on the response messages arriving in time.
	This pattern has also a dual, i.e., One-from-Many Receive-Send, where party $A$ waits for a certain time for messages from $B_1 \dots B_n$, then processes them and returns individual responses.
	
	Related to: scatter-gather.
\end{description}

\subsection{Multi-transmission patterns}
\begin{description}
	\item[{Multi-Responses}.] Party $A$ sends a request to party $B$. $B$ then returns responses until some stop-condition is met. Possible stop-conditions are an explicit notification from $A$, a deadline given by $A$, an interval of inactivity, or an explicit notification from $B$ that the stream has stopped.
	
	Related to: streamed responses, message stream.
	
	\item[{Contingent Requests}.] Party $A$ sends a request message to $B_1$. If there is no response within a certain time, $A$ sends the request to $B_2$, and if there is again a timeout, $A$ continues this cycle until some $B_i$ responds properly.
	
	Related to: send with failover.
	
	\item[{Atomic Multicast Notification}.] Party $A$ sends notifications to parties $B_1 \dots B_n$, where a minimum number of $i$ and a maximum number of $j$ recipients are required to accept the notification.
	A constraint of $i = j = n$ means that all recipients have to accept.
	
	Related to: transactional notification.
\end{description}

\subsection{Routing patterns}

\begin{description}
	\item[{Request with Referral}.] Party $A$ sends a request to party $B$, where the message is evaluated, and based on certain conditions, e.g., message content, follow-up responses are forwarded to a single or multiple parties $C_1 \dots C_n$.
	
	Related to: reply to.
	
	\item[{Relayed Request}.] Party $A$ sends a request to party $B$, and $B$ relays it to parties $C_1 \dots C_n$, who take on further interaction with $A$.
	Party $B$ still retains a view of the ongoing interaction between $A$ and $C_1 \dots C_n$. 
	
	Related to: delegation.
	
	\item[{Dynamic Routing}.] There exist routing conditions that define how a message from party $A$ is forwarded.
	A routing condition can be dynamic, i.e., depend on content.
	Based on the conditions, a request from $A$ is forwarded to one or more parties $B_1 \dots B_n$ that process and eventually forward the message to $C_1 \dots C_n$. These parties then again process, continue to apply  routing conditions and so on.
	
	Related to: routing slip, content-based routing.
\end{description}

\section{Protocols}
\label{sec:protocols}

To deal with the complexity of communication in computer networks, layered protocol design, as proposed by the OSI reference model \cite{Zimmermann1980} or the simplified Internet model \cite{RFC1122}, has become an industry standard to separate concerns in communication protocol design.

\begin{figure}
	\begin{center}
\begin{tikzpicture}[anchor=base]

\node (v1) at (0,-0.4) {\textsf{Link}};
\node (v3) at (0,0.4) {\textsf{Internet}};
\node (v2) at (0,1.2) {\textsf{Transport}};
\node (v4) at (0,2) {\textsf{Application}};

\draw  (-1.2,-0.7) rectangle (1.1,0.1) node (v5) {};
\draw  (-1.2,0.1) rectangle (1.1,0.9) node (v7) {};
\draw  (-1.2,0.9) rectangle (1.1,1.7) node (v9) {};
\draw  (-1.2,1.7) rectangle (1.1,2.5) node (v13) {};
\node at (4.4,-0.8) {IP, IPv6};
\draw  (2,-0.4) node (v8) {} rectangle (6.8,-1);
\node (v6) at (2,-1) {};
\draw  (2,0.7) node (v10) {} rectangle (3.1,-0.4);
\draw  (4,0.7) rectangle (6.8,-0.4);

\draw[dotted]  (v5) edge (v6);
\draw[dotted]  (v7) edge (v8);
\draw[dotted]  (v9) edge (v10);
\draw[]  (2,1.2) node (v11) {} rectangle (6.8,0.7) node (v12) {};
\node at (4.4,0.8) {\emph{Optional Security}: SSL, TLS, DTLS};

\node at (2.5,-0.2) {TCP};
\node at (3.55,-0.2) {SCTP};
\node at (4.6,-0.2) {UDP};
\node at (6,-0.2) {RUDP};
\node at (2.55,0.3) {MPTCP};

\draw[dotted] (v9) edge (v11);
\draw  (2,3.5) node (v14) {} rectangle (v12);

\node at (4.7,0.3) {UDP Lite};
\node at (4.5,4.2) {Content \& Media};

\draw  (6,3.5) rectangle (6.8,1.2);
\node at (6.42,1.9) {DNS};

\node at (3.9,1.4) {FTP};
\node at (3.9,1.9) {SMTP};
\node at (2.6,1.4) {HTTP};
\node at (2.6,1.9) {SPDY};
\draw  (2,2.3) rectangle (3.2,1.2);

\node at (6,0.3) {DCCP};

\node at (5.2,1.4) {XMPP};
\node at (5.2,1.9) {STOMP};
\node at (2.8,2.5) {WebSocket};
\node at (3,3) {WebRTC};
\node at (4.5,3.7) {Data Serialization};
\draw  (2,4.6) node (v15) {} rectangle (6.8,3.5);
\draw[dotted]  (v13) edge (v14);

\node at (5.2,2.4) {AMQP};
\node at (5.2,3) {MQTT};
\node at (3.55,0.3) {QUIC};
\node at (4.2,3) {CoAP};
\node at (4,2.5) {RTPS};
\end{tikzpicture}
	\end{center}
\caption{The Internet model \cite{RFC1122} to the left indicates four conceptual layers, and the right side gives an overview of the discussed protocols}
\label{fig:layers}
\end{figure}
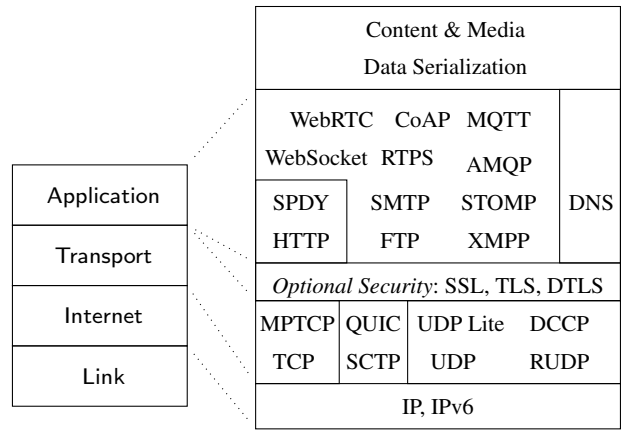

Low delay and multiplexing are two major drivers for recent developments in accelerating Web technology.
Multiplexing in this context refers to techniques for transporting multiple parallel dialogs over a single channel between two peers instead of establishing multiple channels.
This survey approaches the state-of-the-art for bilateral and multilateral information exchange in a bottom-up fashion.
Fig. \ref{fig:layers} recalls the Internet model and a number of communication protocols popular in cloud computing.
Protocols in the \emph{link layer} specify how physically connected devices can exchange information, e.g., IEEE 802.3 Ethernet or IEEE 802.11 Wi-Fi networks.

The \emph{Internet layer} allows hosts to communicate beyond their local neighborhood of physically connected devices.
Logical addresses, routing, and packet-based data exchange are the core aspects of today's Internet.

The \emph{Transport layer} enables inter-process communication over networks.
Multiple processes can run on the same host and share the same logical network address.
Transport layer protocols extend the logical addressing to enable communication between distributed processes.

\emph{Application layer} protocols dictate how to provide functionality, content, and media across two or more processes that are able to communicate, e.g., clients and services.
Such protocols provide transport mechanisms for communicating messages between processes.

While the Internet and Transport layer protocols are typically handled by operating systems, application layer protocols are implemented by the client and service.
From a lexical point of view, all Internet protocols share the same binary alphabet, and a byte is typically the smallest transferable symbol.
A common practice seen in Internet protocols is to separate a transferable sequence of bytes into a protocol header and payload, where the header defines the payload type, so other protocols can be recursively embedded.
Due to the shared alphabet, every protocol needs an unambiguous language to prevent confusion \cite{Sassaman2013}.

\subsection{Internet layer protocols}
\label{sub:connectivity}

The TCP/IP protocol suite \cite{Stevens1993} is the de facto standard for computer networks.
Internet layer functionality is provided by the Internet Protocol (IP) \cite{RFC791}, or IPv4, which defines packet-based networking by an addressing schema, packet layouts, and routing conditions.
Using IP, host $A$ can send a packet to the logical address of host $B$ without knowing the physical address or location of $B$.
Based on the addresses in the header of the packet, so-called routers forward the packet, but delivery is not always guaranteed.
IP therefore implements a Dynamic Routing pattern.
To send and receive packets, a host needs at least one logical IP address in a physically connected network, and a router (gateway) in this network that forwards packets.
If a host is simultaneously connected to two or more physical networks, it is referred to as \emph{multihoming}.

The maximum size of an IP packet is bounded by the underlying link layer technology, and packets are eventually fragmented or dropped if they are too large.
IP Fragmentation allows to split oversized packets into smaller ones, but increases the load on a link because more packets introduce a larger overhead.
The header of an IP packet specifies the type of the enclosed transport protocol in the content.

IP has restrictions; the upper bound of $2^{32}$ logical addresses is the biggest issue.
An ad hoc solution is Network Address Translation (NAT) by routers for private networks.
There is an ongoing effort to switch to IP Version 6 (IPv6) \cite{RFC2460} that provides $2^{128}$ logical addresses to deal with the exhaustion problem that becomes immanent in an Internet of Things.

For addressing multiple recipients with a single packet, IP offers broadcast addresses based on the logical addressing scheme, i.e., a subnet.
IPv6 does not support broadcasts.
Both IP and IPv6 offer multicast, where a packet, sent to a special logical group address, is replicated for all recipients in the group.
Both multicast and broadcast enable the One-to-Many Send pattern on top of Dynamic Routing.

Ideally, the Internet layer promotes \emph{content neutrality}.
All routing decisions should depend on the header of IP packets independent from their content. 
But this neutrality is violated in practice.
Network devices like firewalls, Quality-of-Service traffic shapers, or content-based routers derive routing decisions from payloads of IP packets.
Such devices have functionality across layers in reference models, and they are referred to as \emph{middleboxes} \cite{RFC3234}.
Their techniques are referred to as \emph{Deep Packet Inspection} (DPI) \cite{Bendrath2011}.

%
%
%
%

\subsection{Transport layer protocols}

Transport layer protocols provide content-neutral means of inter-process communication over networks.
Protocols can be distinguished by certain characteristics like:

\begin{itemize}
	\item Uni- or bidirectional communication;
	\item Connection-oriented (stateful) or stateless;
	\item Message- or byte-stream-based information exchange;
	\item Ordering of messages or bytes in a stream;
	\item Reliable delivery;
	\item Data integrity;
	\item Flow and link congestion control.
\end{itemize}

The more properties a protocol supports, the more overhead for control structures is required, and timeliness is affected.
This leads to an upper bound for the maximum transfer rate because there is always a time delay between sending and receiving information.
If, for example, a protocol requires several interactions to synchronize state or acknowledge delivery, the delays accumulate and effectively limit the available transfer rate.

The two most prominent transport layer protocols in the Internet are the Transmission Control Protocol (TCP) \cite{RFC793} and User Datagram Protocol (UDP) \cite{RFC768}.
Both protocols are provided in modern operating systems.
There is an increased interest in enhanced protocols, such as MultiPath TCP (MPTCP)~\cite{RFC6824}, Stream Control Transmission Protocol (SCTP)~\cite{RFC4960}, and Google's Quick UDP Internet Connections (QUIC)~\cite{google-quic} to overcome limitations of TCP and UDP.

\paragraph{Transmission Control Protocol.} \
TCP is a protocol to connect two endpoints, i.e., processes, and information is exchanged in bidirectional byte streams, where the correctness and order of bytes is guaranteed.
For compatibility with packet-based networks, a byte stream is split into TCP segments.
A TCP connection is stateful and establishes a session between the two endpoints.
It requires a so-called \emph{three-way handshake} to synchronize, which causes a delay before the streams can start.
TCP distinguishes a client that initiates the handshake, and a server that listens for incoming connections.
An attempt to reduce the latency caused by the handshake between two already familiar endpoints is TCP Fast Open~\cite{ietf-fastopen}.

The source and destination port number in TCP segment headers identify the client and service endpoints.
There is no payload-type identifier in TCP, and it just transfers byte streams.
IANA \cite{iana} therefore maintains a list of default server listening ports that are automatically assumed by URI schemes if not explicitly overridden, e.g., port \texttt{80/TCP} for HTTP.

Reliable delivery in TCP is achieved by acknowledgments and retransmissions.
Integrity is guaranteed by checksums.
Segment headers also contain a window size that informs the receiver how many bytes the sender can handle in the other directional stream.
The window mechanism enables flow and congestion control through rate adaptation.
The maximum segment size is announced as a TCP option to avoid IP fragmentation.
A problem in TCP is the so-called \emph{head-of-line blocking}; if a single byte in a stream is incorrect or lost, the stream cannot proceed until retransmission succeeds.
This imposes a problem for messaging protocols implemented on top of TCP.
The necessity of acknowledgments limits TCP to bilateral Send and Receive communication.

\paragraph{User Datagram Protocol.} \ 
UDP is a message-oriented unidirectional protocol that transports datagrams without acknowledgments or order.
A datagram header holds a source and destination port, the payload length, and a checksum for integrity, and there is no support for flow and congestion control.
The overhead of UDP is small and it is stateless; therefore, no synchronization is required beforehand.
If the size of a datagram exceeds the maximum payload of the IP packet, fragmentation takes place.
Similar to TCP, a datagram is a sequence of bytes, and the payload type is not identified.
IANA also assigns default server ports to UDP-based protocols.

UDP supports the Send and Receive pattern without delivery guarantees.
Because interaction is stateless, UDP is a candidate for multilateral One-to-Many Send on top of IP broadcast or multicast, where datagrams are replicated by the networking infrastructure.

\paragraph{MultiPath TCP.} \
A limitation of TCP is that segments eventually take the same network path, and a network problem disrupts a connection until error handling routines or timeouts are triggered.
This problem affects mobile devices in radio dead spots or during roaming between wireless networks.
Particularly for mobile devices, it is more and more common that a device is connected to several physical networks simultaneously, i.e., multihoming.
MPTCP is a TCP extension to increase both redundancy and transfer rate by aggregating multiple paths over all available links as subflows of a single connection \cite{Bonaventure2012}.
The MPTCP connection does not fail if a path becomes congested or interrupted and an alternative path is still available.

For compatibility with middleboxes, MPTCP is a host-side extension, and subflows are regular TCP connections. 
A notable example using MTCP is Apple Siri which utilizes both Wi-Fi and 3G/4G networks for increased service availability in mobile devices \cite{apple-mtcp}.
Communication in MPTCP is still bilateral like TCP, i.e., Send and Receive patterns.

\paragraph{Stream Control Transmission Protocol.} \
TCP offers reliable delivery and strict byte order in the stream, but there are applications that require reliability and ordering is less important; TCP can create unnecessary delays in such a scenario \cite{RFC4960}.
Also, the streaming nature of TCP introduces complexity in higher messaging protocols because streams then need a notion of state, delimiters to indicate message boundaries, and measures to circumvent head-of-line blocking.

SCTP \cite{RFC4960} is an alternative to TCP that was designed to raise availability through multihoming as seen in MPTCP.
An association between two endpoints is established in a \emph{four-way handshake}.
A handshake needs more interactions than in TCP, but the SCTP service endpoint stays stateless until synchronization is completed.
This eliminates the well-known security vulnerability of TCP SYN flooding \cite{RFC4987}.

SCTP is message-based and multiplexes byte-streamed messages, similar to MPTCP subflows, in a single association.
SCTP offers reliability through checksums, optional ordering, and rate adaption for flow and congestion control \cite{RFC3286}.
Messages are sequences of bytes, and like TCP and UDP, SCTP does not identify the payload type.
Default listening server ports are managed by IANA.
A drawback of SCTP is its limited popularity: Transportation over the Internet is not guaranteed because middleboxes might block it.
The supported interaction patterns are Send and Receive.

\paragraph{Quick UDP Internet Connections.} \
Developed by Google and already available in the Chrome browser, QUIC \cite{google-quic} is an experimental transport layer protocol to reduce latency and redundant data transmissions in Web application protocols.
QUIC implements the SCTP multiplexing concept on top of UDP to overcome the issue of SCTP being filtered by middleboxes.
Similar to TCP Fast Open, latency is reduced by removing handshakes between already familiar hosts.
QUIC allows transparent data compression, provides checksums and retransmission for reliability, and supports congestion avoidance.
Forward error correction minimizes obstructive retransmissions by an error-correcting code.
In terms of patterns, QUIC offers bilateral Send and Receive.

\paragraph{Other transport protocols.} \
There are several transport layer protocols, where no evident application in a Web or cloud context has been found.
Examples for multilateral interaction are multicast transport protocols \cite{Obraczka1998}.
Examples for bilateral interaction are: the message-based Datagram Congestion Control Protocol (DCCP)~\cite{RFC4340} that offers congestion control, but its delivery is unreliable; UDP Lite~\cite{RFC3828} with relaxed checksums; and Reliable UDP (RUDP)~\cite{ietf-rudp} as an extension of UDP with acknowledgments, retransmissions, and flow control.

\subsection{Transport security}
\label{sub:Encryption}

Several standards for establishing secure sessions, independent from applications, have emerged to achieve confidentiality, integrity, and authenticity in an interaction.
This ``virtual'' security layer is situated between transport and application layer in the Internet model or in the OSI model's session layer.
The most prominent protocols for encryption are the Secure Sockets Layer (SSL) \cite{RFC6101} and Transport Layer Security (TLS) \cite{RFC5246}.
Both operate on top of TCP and establish an encrypted session for byte-stream-based information exchange.
Historically, SSL was invented by Netscape and the latest version 3.0 became obsolete with TLS 1.0 in 1999.
Today's TLS 1.2 still offers limited backward compatibility to SSL.

To establish an encrypted session in SSL/TLS, an initialization routine, i.e., SSL or TLS handshake, is necessary.
The most notable difference between SSL and TLS is \emph{when} the handshake happens.
An SSL handshake takes place implicitly after the TCP handshake, independent from the application.
TLS furthermore allows to trigger a handshake by the application issuing the \texttt{STARTTLS} command to upgrade an already established TCP connection.
During the handshake, the communicating parties authenticate themselves using X.509 public key infrastructure and certificates, they agree on a cipher suite for the session, and securely exchange session keys.
The handshake requires several interactions and adds a delay before encrypted communication can proceed.
For more details, the author refers the reader to Rescola's book \cite{Rescola2001}.

SSL/TLS only works with TCP because the data transfer is assumed to be correct and in order---a handshake or session will fail otherwise.
In MPTCP, subflows are regular TCP connections, and TLS is therefore supported \cite{RFC6897}.
But TLS is not trivial for transport protocols using multiplexing such as SCTP and QUIC.
If a single byte is lost or corrupt, all multiplexed streams will be stalled.
SCTP also supports unordered delivery of messages and partial reliability, which conflicts with the reliability and order assumptions of TLS \cite{RFC3758,RFC6083}.
One approach is to establish a TLS session for every individual stream in the multiplexed connection, but the numerous TLS handshakes accumulate delays and affect timeliness.
TLS is therefore not an optimal choice to secure SCTP and QUIC.

Datagram TLS (DTLS)~\cite{RFC6347} is a modification of TLS that operates on top of UDP.
It specifies a record type as message container, sequence numbers to detect out-of-order records, and retransmission conditions.
DTLS is defined for other transport protocols, e.g., DTLS over DCCP~\cite{RFC5238} and DTLS for SCTP~\cite{RFC6083}.
QUIC specifies its own cryptography protocol loosely based on TLS, and all transmissions are encrypted by default to avoid tampering by middleboxes~\cite{google-quiccrypto}.

\subsection{Application layer protocols}
\label{sub:CommProtocols}

Application layer protocols define inter-process communication on top of transport protocols.
These protocols are often simple service protocols implementing the Send-Receive pattern, where a server waits for incoming communication.
They can be informally distinguished by properties:

\begin{itemize}
	\item Transport layer assumptions;
	\item Text-based or binary protocol;
	\item Stateful or stateless;
	\item Combined or separated data and control connections.
\end{itemize}

Application layer protocols assume properties of transport, e.g., timing, and therefore restrict the allowed transport layer protocols.
Another distinction is whether their syntax is human-readable, and text-based protocols typically have an overhead because reduced information density.
A protocol is said to be stateful if the result of a previous interaction affects the choice of the next interaction.
Furthermore, protocols can either mix control and content in a single connection or maintain separate connections.

It should be noted that some protocols distinguish between a conceptual high-level syntax and a wire format that is actually transmitted; e.g., while high-level syntax could be text-based, the effectively sent data could be compressed.

Two well-known text-based application layer protocols are the File Transfer Protocol (FTP) \cite{RFC959} and the Simple Mail Transfer Protocol (SMTP) \cite{RFC5321}.
Both require TCP and support TLS for security; while FTP maintains two separate TCP connections for text-based control commands and binary media transfer, SMTP sends the text-based control commands and the email message in a single TCP connection.
Both protocols are stateful because they require several Send-Receive interactions, where success or failure of the current interaction decides the next action.

\subsubsection{Domain name system}
\label{sub:dns}

The Domain Name System (DNS) \cite{RFC1034,RFC1035} is an Internet core service, managed by IANA, and specifies an application layer protocol.
As names are more usable for humans than numeric addresses, DNS is a distributed database for a hierarchical naming scheme that maps names onto IP addresses.
Records in DNS have a certain type; e.g., type \texttt{A} is a host address record, type \texttt{CNAME} is an alias for another name, or type \texttt{MX} is reserved for SMTP-service-specific records.
The hierarchical name of a host is then referred to as Fully Qualified Domain Name (FQDN).

DNS is a binary and stateless protocol implementing the Send-Receive pattern.
A client queries a service to resolve a name of a certain type, and the service eventually returns a record.
DNS uses UDP as transport protocol for queries but also supports TCP for large responses or DNS transactions, i.e., zone transfers.
The drawback of using TCP for short queries is the handshake delay.

DNS has become a critical service for today's Internet, and the majority of services relies on DNS as an abstraction layer for locating endpoints; DNS service records (type \texttt{SRV}) even enable dynamic service discovery~\cite{RFC6763}.
Nevertheless, a failure, misuse, or misconfiguration in DNS can lead to unforeseeable security consequences; e.g., attacks like DNS spoofing \cite{Herzberg2013} are a serious threat.
If DNS responses are tampered with, an attacker can redirect interaction to malign hosts.
DNSSEC \cite{RFC4033,RFC4035,RFC4034} is therefore an attempt to secure correctness and authenticity of queries and responses using cryptographic methods.

\subsubsection{Resource identification and location}
\label{sub:uri}

A Uniform Resource Identifier (URI) is a human-readable text string that identifies a resource in the Web.
The URI specification \cite{RFC3986} defines the syntax of a URI and also describes a Uniform Resource Locator (URL), a subtype of URI that both identifies and locates a resource.

\begin{figure}
	\begin{center}
\usetikzlibrary{decorations.pathreplacing}
\begin{tikzpicture}[anchor=base]
\node at (-2,0) {\texttt{http://www.ex.com/dir/fotos.php?action=view\#page2}};
\draw [decorate,decoration={brace,amplitude=5pt},xshift=-4pt,yshift=0pt]
(-4.95,-0.1) -- (-5.75,-0.1) node [black,below,midway,yshift=-2mm] {scheme};
\draw [decorate,decoration={brace,amplitude=5pt},xshift=-4pt,yshift=0pt]
(-3,-0.1) -- (-4.7,-0.1) node [black,below,midway,yshift=-2mm] {authority};
\draw [decorate,decoration={brace,amplitude=5pt},xshift=-4pt,yshift=0pt]
(-0.8,-0.1) -- (-2.9,-0.1) node [black,below,midway,yshift=-2mm] {path (\emph{identity})};
\draw [decorate,decoration={brace,amplitude=5pt},xshift=-4pt,yshift=0pt]
(1.1,-0.1) -- (-0.7,-0.1) node [black,below,midway,yshift=-2.7mm] {query};
\draw [decorate,decoration={brace,amplitude=5pt},xshift=-4pt,yshift=0pt]
(2.0,-0.1) -- (1.2,-0.1) node [black,below,midway,yshift=-2mm] {fragment};
\draw [decorate,decoration={brace,amplitude=5pt},xshift=-4pt,yshift=0pt]
(-4.7,0.3) -- (-0.8,0.3) node [black,above,midway,yshift=2mm] {hierarchical part (\emph{location})};
\end{tikzpicture}
	\end{center}
\caption{An example URL identifies and locates a resource.}
\label{fig:url}
\end{figure}
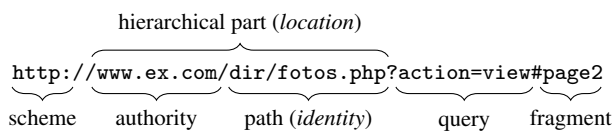

Fig. \ref{fig:url} shows an example.
The URI scheme identifies the application layer protocol for accessing the resource and implicitly assumes the default TCP or UDP service listening port from a database maintained by IANA, e.g., \texttt{ftp:} for FTP or \texttt{mailto:} for SMTP.
The authority contains a Fully Qualified Host Name (FQHN) which is either a FQDN or an IP address to locate the host of the service.
Also, additional user credentials and alternating TCP or UDP ports can refine the authority.
The path identifies a resource within the authority, an optional query part holds a list of key-value pairs, and an optional fragment can refer to specific information within the resource.

A URL has a notion of \emph{origin}. Informally, two URLs share the same origin if they have identical scheme, FQHN, and port \cite{RFC6454}.
As URLs have a fundamental role in the Web, there is a trend to simplify URLs to improve usability.
So-called \emph{clean URLs} \cite{Opitz2006} have an self-explanatory path and avoid the query part, e.g., the clean URL for Fig.~\ref{fig:url} is then \nolinkurl{http://www.ex.com/fotos/view\#page2}.
The rewriting of URLs into a clean form, automatically and manually, is a common practice in Web development.

\subsubsection{The hypertext transfer protocol}
\label{sub:HTTP}

\begin{figure*}
	\begin{center}
\usetikzlibrary{decorations.pathreplacing}
\begin{tikzpicture}
\node[align=left, draw,font=\ttfamily\scriptsize] (req) at (-8.1,-2.3) {GET /dir/fotos.php?action=view HTTP/1.1\\
Host: www.ex.com\\
Connection: keep-alive\\
Accept: */*\\
User-Agent: Mozilla 5.0\\
Accept-Encoding: gzip,deflate,sdch\\
Accept-Language: en-US,en;q=0.8,de;q=0.6};

\node[align=left,draw,font=\ttfamily\scriptsize] (res) at (-1.3,-2.3) {HTTP/1.1 200 OK\\
Date: Mon, 02 June 2014 10:15:22 GMT\\
Server: Apache\\
Last-Modified: Mon, 01 June 2013 00:00:00 GMT\\
Accept-Ranges: bytes\\
Content-Length: 431\\
Cache-Control: max-age=0, no-cache, must-revalidate\\
Expires: Mon, 02 June 2014 10:15:22 GMT\\
Etag: "4135cda4"\\
Connection: keep-alive\\
Content-Type: text-html\\
\\
<!DOCTYPE html>\\
<html xml:lang="en" lang="en"....};

\node (v1) at (-5.3,0.2) {\textsf{Service}};
\node (v2) at (-5.3,-4.8) {\textsf{Client}};
\draw[->, >=latex]  (req) |- (v1);
\draw[->, >=latex]  (res) |- (v2);
\draw[->, >=latex]  (v2) -| node[left] {\textbf{1. HTTP Request}} (req);
\draw[->, >=latex]  (v1) -| node[right] {\textbf{2. HTTP Response}} (res);

\draw [decorate,decoration={brace,amplitude=3pt},xshift=-4pt,yshift=0pt]
(-10.8,-1.55) -- (-10.8,-1.3) node [black,left,midway,xshift=-1mm] {Method, path};
\draw [decorate,decoration={brace,amplitude=3pt},xshift=-4pt,yshift=0pt]
(-10.8,-3.3) -- (-10.8,-1.65) node [black,left,midway,xshift=-1mm] {Header fields};
\draw [decorate,decoration={brace,amplitude=3pt},xshift=-4pt,yshift=0pt]
(2.4,-0.3) -- (2.4,-0.6) node [black,right,midway,xshift=1mm] {Status};
\draw [decorate,decoration={brace,amplitude=3pt},xshift=-4pt,yshift=0pt]
(2.4,-0.7) -- (2.4,-3.4) node [black,right,midway,xshift=1mm] {Header fields};
\draw [decorate,decoration={brace,amplitude=3pt},xshift=-4pt,yshift=0pt]
(2.4,-3.7) -- (2.4,-4.3) node [black,right,midway,xshift=1mm] {Content};
\end{tikzpicture}
	\end{center}
\caption{After the client has established a TCP connection with the service, a request is sent as text-based stream. The service parses the request and returns a response containing a header and the resource through the other directional stream of the TCP connection. Depending on the HTTP method, a request eventually has a content, e.g., from a form submission or file upload.}
\label{fig:http_reqres}
\end{figure*}
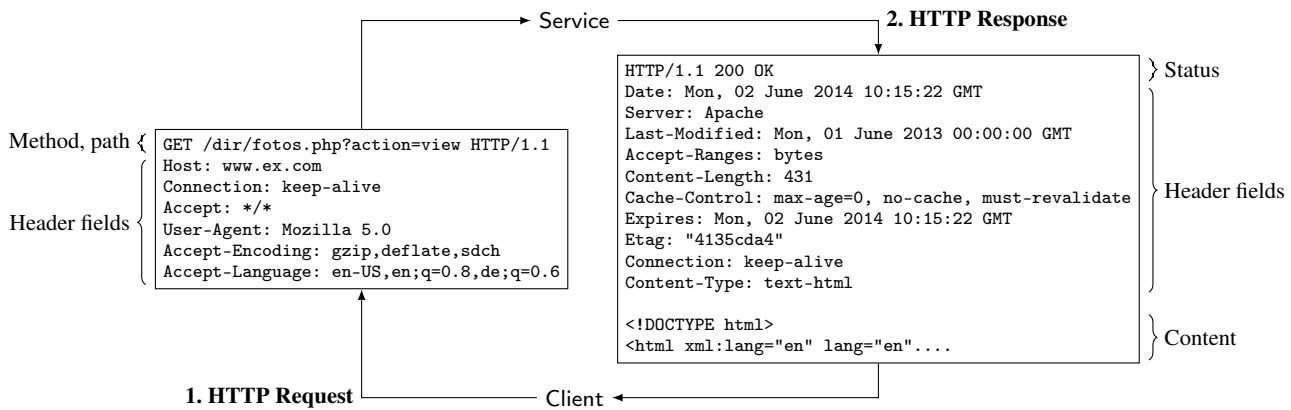

HTTP, originally specified in RFC 2616 \cite{RFC2616}, is the fundamental application layer protocol for the Web and many service technologies.
It is stateless and implements the Send-Receive pattern: A client sends a request message, and the service answers with a response message.
HTTP is designed to operate on top of TCP, and both control and content are sent in a single TCP connection, where control instructions are text based.
To separate control from data, HTTP specifies a header format and delimiters.
Fig. \ref{fig:http_reqres} shows a Send-Receive cycle between a client and a service.

Both HTTP request and response messages specify a header and an optional body separated by a delimiter.
The request header defines the method, the URI-path of a resource, the protocol version, and a list of header fields.
The presence of a body depends on the method.
Similarly, the response header holds a status code, a list of response header fields, and eventually a body, i.e., the requested content.

Header fields make HTTP extensible.
Some headers are mandatory, others are optional.
\texttt{Content-Type} is a central header field to specify the MIME type of media.
The content type of a requested resource is therefore undefined until the HTTP response message arrives at the client.

Today's HTTP/1.1 optimizes its predecessor versions in several ways.
HTTP/1.1 supports eight methods: \texttt{OPTIONS}, \texttt{GET}, \texttt{HEAD}, \texttt{POST}, \texttt{PUT}, \texttt{DELETE}, \texttt{TRACE}, and \texttt{CONNECT}.
The \texttt{Upgrade} header enables the client and service to switch the application protocol after a Send-Receive cycle.
The \texttt{Host} header distinguishes FQDN authorities that are hosted on the same server---a common practice in Web hosting.

Also, HTTP/1.1 introduces persistent TCP connections and pipelining of requests to minimize the accumulating delay caused by the TCP handshakes for every requested resource.
The standard proclaims that a client should not exceed two simultaneous TCP connections to a service.
HTTP differentiates \texttt{Content-Encoding} of the requested resource and \texttt{Transfer-Encoding} for transport-only compression between client and service.
HTTP also offers fine-grained caching mechanisms, e.g., by timestamps, the \texttt{ETag} header, and conditional \texttt{GET}, such that clients and so-called proxies can minimize data transfer.

HTTP/1.1 allows client- or service-driven content negotiation for resources \cite{RFC7231}.
In service-driven negotiation, the service considers the client's \texttt{User-Agent}, client-side content-type restrictions (\texttt{Accept}), accepted character encodings for text-based formats (\texttt{Accept-Charset}), accepted content encodings (\texttt{Accept-Encoding}), and personal natural language preferences (\texttt{Accept-Language}) for sending a suitable representation.
In client-driven negotiation, also referred to as agent driven, the service returns a multiple-choices status in a first response that lists all available representations, where the client can choose from in a second Send-Receive cycle.

While HTTP is stateless in principle, HTTP/1.1 adds support for so-called \emph{cookies} using the \texttt{Set-Cookie} and \texttt{Cookie} header fields to track state across Send-Receive cycles \cite{RFC6265}.
A cookie is basically a text string that identifies a client's HTTP session.
The responsibility for correct application state tracking is on the service side; cookies therefore have important security and privacy aspects.

Recently, the HTTP/1.1 standard has been completely respecified in order to remove imprecisions that have led to ambiguous implementations \cite{RFC7233,RFC7234,RFC7235,RFC7232,RFC7230,RFC7231}.

\paragraph{HTTP security.} \
The use of SSL/TLS for TCP connections has become the de facto standard to secure HTTP interaction between a client and a service, and it is specified as HTTP Secure (HTTPS) \cite{RFC2818}.
As HTTPS has its own URI scheme (\texttt{https:}), a user can recognize from a URL whether the access to a resource is protected.

A client can eventually choose to access an identical resource through multiple transport mechanisms, e.g., HTTP or HTTPS.
To notify a Web client that resources of a certain authority can only be accessed through HTTPS, the HTTP Strict Transport Security (HSTS) \cite{RFC6797} specifies a response header field that informs the client about this policy.

Using the \texttt{CONNECT} HTTP method, a client can ask a proxy service to establish a connection to the intended service on behalf of the client, and byte streams are forwarded.
This functionality, referred to as HTTP tunneling, is required in proxies for HTTPS access to services.
As exchanged data are encrypted, caching is not possible.

Another way to secure HTTP interaction is to upgrade an existing TCP connection to a TLS session by using the \texttt{Upgrade} header in HTTP/1.1 \cite{RFC2817}.
A drawback of this solution is that a user can no longer see from a URL whether access is encrypted or not.

\paragraph{Push technology.} \
In terms of patterns, a Send-Receive interaction in HTTP is synchronous and can only be initiated by the client; resources are \emph{pulled} from a service.
Due to wide availability of HTTP-enabled software and acceptance by middleboxes, HTTP has been exploited to achieve asynchronous interaction without breaking the protocol specification, e.g., for client-side Receive or Multi-Responses patterns.
These techniques are commonly referred to as push technology~\cite{Alinone2011}, so a service can \emph{push} a resource to a client preferably in real time, e.g., for data feeds or event notification, without being explicitly requested.
Historically, the first attempts have resorted to client-side polling, and real-time event notification was not possible.
To minimize the number of Send-Receive cycles and to decrease response times, long polling is similar to polling, but the HTTP request \emph{hangs}, i.e., is not answered, until a server-side event or a timeout occurs.

Comet \cite{Russell2006}, also known as HTTP Streaming or HTTP server push, exploits persistent connections in HTTP/1.1 to keep a single TCP connection open after a client requests an event resource.
The service then gradually delivers events using MIME type \texttt{multipart/x-mixed-replace} for the response.
Comet implements the Multi-Responses pattern.

Reverse HTTP \cite{ietf-reversehttp,reversehttp,Sit2000} exploits the \texttt{Upgrade} feature of HTTP/1.1 to change the application layer protocol and switch the roles of client and service.
The service becomes an HTTP client in the established TCP connection, and real-time events are then issued as HTTP requests from the original service to the original client, i.e., client-side Receive-Send in terms of patterns.

For simultaneous bidirectional communication between client and service, Bidirectional-streams Over Synchronous HTTP (BOSH) \cite{xep-bosh} maintains two separate TCP connections.
The client uses the first connection to issue HTTP request messages to the service, the second connection is a hanging request initiated by the client, so the service can interact with the client asynchronously.
This enables patterns Send and Receive for both client and service.

Two recent Web techniques in HTML5 are Server-Sent Events (SSE)~\cite{w3c-sse} and WebSocket~\cite{RFC6455,w3c-websocket}.
For SSE, the client requests a resource that acts as event resource similar to Comet, i.e., implements the Multi-Responses pattern.
The response is of MIME type \texttt{text/event-stream}, the TCP connection is kept open, and events are delivered as byte chunks.
In case of a timeout, the client reconnects to the event resource.

WebSocket establishes a bidirectional channel for simultaneous communication in both directions.
A WebSocket connection behaves like a bidirectional byte-stream-oriented TCP connection between client and service for Send and Receive interaction, and it is established in a handshake by exploiting the HTTP \texttt{Upgrade} header in HTTP/1.1.
During this HTTP Send-Receive cycle for the WebSocket handshake, properties are negotiated using HTTP headers, including \texttt{Sec-WebSocket-Protocol} to agree on a subprotocol for continuation after the handshake.
WebSocket supports operation on top of TLS and provides individual URI schemes for unencrypted (\texttt{ws:}) and encrypted (\texttt{wss:}) communication.

With respect to Comet, Reverse HTTP, or BOSH, WebSocket has the smallest overhead because it is independent of HTTP when established.
Nevertheless, clients and services need an explicit application layer protocol to operate on top of a WebSocket connection.

\paragraph{Performance and speed.} \
Performance is an issue in HTTP when many resources are requested simultaneously.
Even when HTTP/1.1 persistent connections and pipelining are supported, access becomes somehow serialized because of limited simultaneous TCP connections.
Allowing more parallel connections has a negative effect on the availability of a service because there is an upper limit, how many TCP connections a host can serve simultaneously.
A workaround for the connection limit is \emph{domain sharding} \cite{Souders2009}; resources are distributed over multiple authorities, controlled by the service provider, so a client can use the maximum number of TCP connections to every authority in parallel.
In general, there are a several approaches to increase Web performance and user experience: 

\begin{itemize}
	\item minimize protocol handshake latency;
	\item reduce protocol overhead;
	\item multiplexed access;
	\item prioritization of access.
\end{itemize}

Two standards that have never left the experimental status are the binary HTTP-NG \cite{w3c-httpng}, intended as a successor of HTTP, and multiplexing HTTP access over a single TCP connection based on SMUX \cite{w3c-smux}.
Other experimental multiplexing approaches are Structured Stream Transport~\cite{Ford2007} and HTTP over SCTP \cite{Natarajan2008}.

The state-of-the-art, Google SPDY \cite{google-spdy}, acts conceptually as a session layer protocol between TCP and HTTP to increase Web performance through multiplexed resource access, prioritization, and compression of headers and content to reduce overhead.
SPDY changes the wire format, but retains the semantics of HTTP; it is basically an augmentation to HTTP and no individual URI scheme is specified for compatibility reasons.
SPDY also allows service-initiated interaction to push related resources to the client before they are asked for, i.e., the Multi-Responses interaction pattern.

As SPDY changes the wire format of HTTP, encryption is mandatory to prevent middleboxes from tampering with interactions.
SPDY requires TLS with Next Protocol Negotiation (NPN) support for backward compatibility with HTTPS.
When a client accesses an HTTPS service, the service announces SPDY support through NPN during the TLS handshake, and the client can choose to proceed with SPDY or traditional HTTP within the TLS session.
The experimental QUIC transport protocol was specifically designed for SPDY to remove delays between familiar hosts, caused by the initial TCP handshake, and optimized flow control~\cite{google-quic}.

SPDY and WebSocket have been proposed as two core technologies in the upcoming HTTP/2 which is currently in the specification process \cite{ms-http20}.
SPDY is already confirmed as the basis for HTTP/2, but protocol negotiation will be switched from NPN to the more general TLS Application Layer Protocol Negotiation (APLN) mechanism in the future \cite{http20}.
Contrary to SPDY, encryption in HTTP/2 is not mandatory; some implementations have stated that they will only support HTTP/2 over an encrypted connection \cite{ietf-http2}.

\subsubsection{Web client-to-client communication}
\label{sub:Client2ClientComm}

A recent development on the client side, already supported by many modern browsers, is Web Real-Time Communications (WebRTC) \cite{google-webrtc,w3c-webrtc} to enable direct interaction between clients.
The motivation is real-time information exchange without the necessity of a third-party service or broker for video and voice calls to avoid network delays and bottlenecks.

WebRTC uses UDP as transport and still needs a service for discovery and signaling between clients, but also for dealing with NAT or firewalls on the path between two clients.
WebRTC is not limited to audio and video streaming.
For general information exchange, WebRTC features a so-called \texttt{RTCDataChannel} to establish a direct SCTP association, protected by DTLS, between two clients \cite{Ristic2014}.
Interaction in WebRTC is message based and an association can provide ordering and reliable transfer.
In terms of patterns, an \texttt{RTCDataChannel} supports Send and Receive interaction between clients.

\subsubsection{Messaging protocols}
\label{sub:MessageingProtocol}

A increasingly popular group of protocols is designed for messaging passing, where peers interact through a message-oriented middleware or a message broker service.
Messaging solutions often specify their own wire formats, i.e., application layer protocols, and this subsection enumerates the most important ones with respect to cloud computing.
Architectures utilizing these protocols are then discussed in Sect. \ref{sub:mom}.

\paragraph{Proprietary wire formats.} \ The Microsoft Message Queuing (MSMQ) \cite{ms-msmq} service specifies individual wire formats and utilizes TCP and UDP during interaction.
Version 3.0 of MSMQ also introduces messaging through HTTP or HTTPS to overcome middleboxes.
For sending messages to multiple recipients on different hosts, MSMQ supports IP multicast, so message replication is implicitly performed by the network, and not MSMQ.

TIBCO offers several proprietary messaging solutions such as the Enterprise Message Service (EMS) \cite{tibco-ems} or Rendezvous \cite{tibco-rendezvous} for enterprise and cloud messaging.
While EMS utilizes individual XML-based message formats sent over TCP or WebSocket \cite{tibco-wm}, Rendezvous uses a proprietary binary message format sent over UDP and IP multicast for One-to-Many Send interaction.

The OpenMQ binary wire format is a protocol for Java Glassfish \cite{openmq}.
Another individual binary format is OpenWire \cite{apache-openwire} for Apache ActiveMQ \cite{apache-activemq}.
Apache Kafka \cite{apache-kafka} also specifies a binary protocol on top of TCP transportation.
ZeroMQ \cite{imatix-zeromq} is an intelligent socket library for exchanging arbitrary binary messages, and ZeroMQ specifies a protocol to split those messages into one or more frames for transportation over TCP.

\paragraph{Standardized wire formats.} \
Today's most prominent open standard for messaging is the Advanced Message Queuing Protocol (AMQP) \cite{Vinoski2006}.
The AMQP 1.0 transport model is an OASIS standard \cite{oasis-amqp} that recently became the international standard ISO/IEC 19464 \cite{iso-amqp}.
The transport model specifies a binary wire format that multiplexes channels into a single TCP connection or SCTP association.
It supports flow control for messaging, authentication \cite{RFC4422}, and encryption by TLS.
AMQP is stateful because communicating peers negotiate a session with a handshake after an TCP connection has been established.
Peers exchange so-called frames that contain header fields and binary content.
For interoperable representation of messages, AMQP offers a self-contained type system that includes a set of primitive datatypes, descriptors for specifying custom types, restricted datatypes, and composite types for structured information.
This allows self-describing annotated content when interaction between heterogeneous platforms takes place.

Known as Jabber, and initially motivated by portable instant messaging, the Extensible Messaging and Presence Protocol (XMPP)~\cite{RFC6120,RFC6121} is another standard for messaging.
Short text-based messages, i.e., XML stanzas, are bidirectionally exchanged in open-ended XML streams over a long-lived TCP connection, eventually protected by TLS, between a client and service or service-to-service.
To overcome middleboxes, XMPP can utilize HTTP and push technology, i.e., BOSH \cite{xep-boshxmpp}.
Also, XMPP over WebSocket is in an experimental state \cite{ietf-xmppws}. 

The text-based Streaming Text Oriented Messaging Protocol (STOMP) \cite{stomp} is also an interoperable messaging protocol that has resemblance to HTTP.
It operates on top of a bidirectional byte-stream-based transport protocol such as TCP or WebSocket, supports TLS for encryption, and uses UTF-8 as default character encoding.

MQ Telemetry Transport (MQTT) \cite{ibm-mqtt} is an open standard for lightweight messaging on top of TCP and low bandwidths as encountered in the Internet of Things.
It defines a binary message format with a small fixed-size header of only two bytes and therefore little overhead.
MQTT also supports SSL/TLS for encrypted transfers.

The Constrained Application Protocol (CoAP) \cite{RFC7252} is another standard for the Internet of Things, where hardware is typically constrained.
CoAP has a binary message format for asynchronous messaging over UDP.
The standard specifies mappings between CoAP and HTTP, and they have similar semantics.
CoAP offers optional delivery guarantees using acknowledgments, supports Send-Receive, and also allows multilateral interaction using IP Multicast.
The standard refers to DTLS for securing CoAP interactions.

The Data Distribution Service for Real-Time Systems (DDS) \cite{omg-dds} is a machine-to-machine middleware specification with applications in the Internet of Things.
DDS also specifies the Real-Time Publish-Subscribe (RTPS) \cite{omg-ddsi} binary wire protocol for TCP- and UDP-based messaging, including IP multicast for One-to-many Send interaction.

\section{Architectures}
\label{sec:architectures}

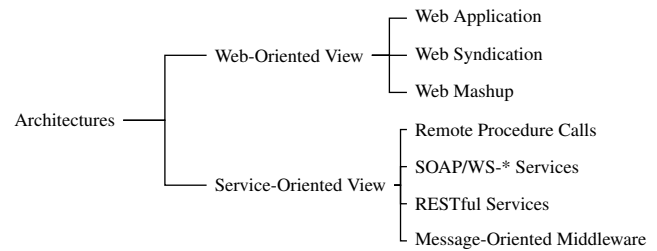
\begin{figure}
	\begin{center}
\begin{tikzpicture}[level distance=75pt, font=\scriptsize]
\tikzset{level 1/.style={level distance=65pt}}
\Tree[.{Architectures}
	[.{Web-Oriented View}
		{Web Application}
		{Web Syndication}
		{Web Mashup}
	]
	[.{Service-Oriented View}
		{Remote Procedure Calls}
		{SOAP/WS-* Services}
		{RESTful Services}
		{Message-Oriented Middleware}
	]
]
\end{tikzpicture}
	\end{center}
\caption{The survey distinguishes a Web- and service-oriented view on architectures}
\label{fig:architectures_overview}
\end{figure}

An architecture combines protocols, languages, and service interaction patterns for service delivery.
Fig. \ref{fig:architectures_overview} describes the structure of this section.
Architectures are distinguished into a Web- and service-oriented view based on their evolution: While the Web-oriented view explores typical Web scenarios, the service-oriented view focuses on well-known architectures from enterprise service integration and cloud computing.

\subsection{Web-oriented view}

The World Wide Web is all about hypermedia: To present multimedia content like text, audio, and video in a nonlinear fashion, where users can access information through hyperlinks.
The Web has evolved through several phases that are typically referred to by \emph{Web 1.0, 2.0,} and \emph{3.0} \cite{Endres-Niggemeyer2013}.

While \emph{Web 1.0} refers to the advent of accessible hypermedia, contents were often static, non-interactive and created only by few \cite{Cormode2008}.
The key aspects of \emph{Web 2.0} are technological advances for dynamic content, interactive user interfaces, and also a social community aspect; users can interactively collaborate and create new hypermedia content.
Some examples are blogs, electronic marketplaces, or social networking.
\emph{Web 3.0} is believed to add personalization based on semantics of content, for example, adaptivity to user preferences and context.
The user experience in discovering knowledge is expected to go beyond merely following hyperlinks and include social, mobile, and location aspects of the user.

Web applications are discussed as the reference architecture for hypermedia delivery, while Web syndication and mash\-ups are a form of composition.

\subsubsection{Web application}

A Web application, hosted on a webserver, delivers websites to clients.
The components of a website are HTML/XHTML markup, CSS for visual representation, JavaScript code, and media resources.
Resources are uniquely identified by URLs and a client, i.e., a Web browser or user agent, requests them from a Web application using HTTP or HTTPS as transport mechanism:

\begin{itemize}
	\item A website structures text content and media resources using HTML or XHTML markup.
	A client retrieves the markup and parses it into a DOM tree.
	\item The markup can define or refer to a CSS resource that specifies the visual representation of the DOM, e.g., colors, fonts, images, or effects.
	\item The JavaScript runtime environment executes nested or embedded script code in context of the DOM to interactively update the DOM tree during runtime.
	\item The DOM is eventually rendered in a window.
\end{itemize}

For composing multimedia and text, tags in the markup can refer to other resources via URLs.
Depending on the content type of the resource, either predefined in the markup or announced in HTTP, the client either parses and presents the resource natively or hands it to an appropriate \emph{plugin} during runtime.
The HTTP optimizations discussed in Sect. \ref{sub:HTTP} aim to minimize the delay experienced by the user when multiple resources need to be loaded to compose a website.

After loading, the user has a choice of nonlinear continuations through hyperlinks.
A hyperlink refers to a URL, and when followed, a full page transition is triggered; the current DOM is replaced, and embedded resources are recursively loaded again.
As common in Web 1.0, if a resource changes on the service-side, the client needs to poll to recognize changes, construct a completely new DOM and load all uncached resources.

For stateful page transitions, Web applications resort a unique identifier in the URL, in hidden form fields, or use a cookie in HTTP to associate a client request to a service-side session to track application state.
If a cookie has been set for a particular origin, they are automatically added to HTTP request headers.
This allows, for example, an authenticated session spanning over several page transitions.

In terms of interaction patterns, a traditional Web application inherits bilateral Send-Receive from its application layer protocol HTTP or Multi-Responses in case of audio or video streaming over HTTP.

\paragraph{Browser content policies.} \ The \emph{same-origin policy} on the client side is the fundamental security model of Web applications.
In general, the DOMs of markup documents retrieved from different origins are isolated from each other for security reasons, including access and transmission of cookies.
The same-origin policy applies to network-accessible resources and script APIs \cite{mdn-sameorigin}.
While the policy allows outgoing write access, e.g., form submissions, and static embedding of foreign resources in markup, dynamic read access during execution of script code is only allowed to resources from the same origin.
Access to a DOM and its APIs through script code is also restricted.
Script code is executed in context of the DOM that embeds the script code.
If two DOMs that share the same origin execute two scripts simultaneously, both scripts can fully access each other's DOMs and APIs.
The same-origin policy enforces a coarse access control in Web browsers to isolate and separate interaction between simultaneously active websites.

The same-origin policy prevents Web applications from using script code to dynamically compose resources from different origins.
A recent enhancement in browser security for a more fine-grained access control to foreign resources is Content Security Policy (CSP) \cite{w3c-csp}.
CSP specifies additional HTTP headers and content type families.
A service can then notify a client about trusted third-party origins with respect to certain content type families of resources.

\paragraph{Distributed authoring and versioning.} \
One of the earliest standards for user collaboration over the Web is the Web Distributed Authoring and Versioning (WebDAV) \cite{RFC4918} extension for HTTP.
While the first evolution of the Web has offered read-only content for most of the users, WebDAV adds write access such that users can edit resources together.
This is accomplished by adding new methods and header fields to the HTTP/1.1 standard; therefore, both client and service have to implement this extension.
Two widely used extensions of WebDAV are CalDAV~\cite{RFC4791} for shared calenders and task lists and CardDAV~\cite{RFC6352} for sharing address books.

\paragraph{Asynchronous JavaScript and XML.} \
A characteristic development of Web 2.0 is toward dynamic Web applications and improved user experience.
Instead of full page transitions, only parts in a DOM are updated during runtime using Asynchronous JavaScript and XML (AJAX)~\cite{Garrett2005}.
AJAX relies on the client-side \texttt{XMLHttpRequest} API to initiate HTTP or HTTPS interaction from script code.

Instead of delivering a large HTML markup content that requires substantial time for parsing and loading of embedded objects, an AJAX-enabled Web application serves only a small markup skeleton and script code, so the client requests the slower-loading objects asynchronously and updates the DOM when available.
AJAX updates can also be triggered by client-side events, e.g., user interface events.
AJAX explicitly refers to XML as format for updates, but the mechanism can in fact accept any text-based content type, e.g., JSON, and process it by script code during runtime.

HTTP push technology, as discussed in Sect. \ref{sub:HTTP}, e.g., Comet or WebSocket, enables asynchronous near-real-time updates from the service to the client in collaboration with AJAX.
User-experienced page load times and interactivity therefore improve.

The same-origin policy applies to AJAX, i.e., accessing foreign resources through an \texttt{XMLHttpRequest} is considered as a read access and only allowed to resources that share the same origin as the executed script code.
Cross-Origin Resource Sharing (CORS)~\cite{w3c-cors} allows \texttt{XMLHttpRequests} to a foreign resource, but this access has to be explicitly approved by the foreign Web application using HTTP headers.

For cross-origin access to JSON resources, when a client does not support CORS, there exists a workaround by script embedding, called JSON with Padding (JSONP) \cite{jsonp}.
The same-origin policy does not apply to resources statically embedded in markup.
A JSON document is valid JavaScript code, and JSONP exploits the \texttt{script} tag to load the JSON document padded by script code from a specific URL.
The padded script code then calls a user-defined function on the client side to process the JSON object.

\paragraph{Semantic Web.} \ Information in the Web is primarily encoded in semi-structured HTML or XHTML documents, often using natural language, which turns discovering, sharing, and combining information into a hard problem.
The Semantic Web is an attempt to standardize formats, so the ``Web of documents'' can become a machine-interpretable ``Web of linked data'' \cite{w3c-semanticweb}.
Linked data are believed to be a major characteristic of the Web 3.0.

The W3C-proposed Semantic Web has three technological pillars: XML and Unicode as fundamental language; the Web Ontology Language (OWL)~\cite{w3c-owl} to express relations; and the Resource Description Framework (RDF)~\cite{w3c-rdf} as a collection of specifications including vocabularies, a metadata data model, and serialization formats.

Several standards have been proposed to increase applicability of W3C's Semantic Web in today's document-based Web.
RDF through attributes (RDFa) \cite{w3c-rdfa} is an extension to HTML and XHTML, where semantics of existing markup are annotated using HTML or XHTML attributes.
Also, Microdata~\cite{w3c-microdata} for HTML5-based documents and JSON for Linking Data (JSON-LD)~\cite{w3c-jsonld} are annotation based and compatible with RDF.
Microformats~\cite{microformats} is different annotation-based approach, independent from W3C Semantic Web specifications.

\paragraph{HTML5 and assisting technologies.} \ The numerous versions of HTML, XHTML, and ambiguities in their specifications have led to incompatible implementations or inconsistent visual representations in Web browsers.
Rich Web client functionality has relied on plugins, e.g., Adobe Flash, which had a negative impact on the accessibility for many clients when the plugin was not supported.

Today, HTML5 is an attempt to define an unambiguous syntax for a unified reference markup language and also to standardize APIs for a number of assisting technologies that have contributed to the success of Web 2.0 in building rich Web applications \cite{w3c-html5}, such as:

\begin{itemize}
\item Semantic annotation of markup through Microdata;
\item A set of natively supported audio and video formats;
\item App Cache for offline storage of HTML5 websites;
\item File System API for persistent storage;
\item Web Storage and IndexedDB for temporary and offline storage of client data;
\item Concurrent execution of script code by Web Workers;
\item Web Messaging, also called cross-document messaging, for controlled data exchange between DOMs of different origin;
\item Server-Sent Events (SSE) and WebSocket to increase interactivity of websites that require high responsiveness;
\item Geolocation access for localized personalization.
\end{itemize}

There are also non-W3C technologies to improve capabilities of the client, in particular, WebGL~\cite{khronos-webgl} for 3D rendering support, and WebCL~\cite{khronos-webcl} for exploiting parallel computing hardware on the client.
Also, the WebRTC API is available in many modern browsers today, but not part of the HTML5 specification yet.
HTML5 is an evolutionary step toward accessible websites and rich Web client experience across Web browsers and mobile devices.

\subsubsection{Web syndication}
\label{sub:WebSyndication}

While a Web application is a service for bilateral interaction with a client, the client needs to poll or rely on HTTP push technology to recognize service-side changes.
Web syndication propagates event notifications or push content from services to clients and also to other Web applications for service-to-service interaction, e.g., blogs.

The simplest form of change notification is a service-side webhook \cite{Lindsay2007}.
There are two parties: a peer that experiences an event and triggers an HTTP request to a webhook URL, and a routine that handles requests and processes posted information for the webhook URL.
The definition of a webhook is kept abstract, and there is only a trigger and a handler in the spirit of callback or event programming.
Lindsay \cite{Lindsay2012} further argues that callback principles in the Web will eventually lead to the \emph{Evented Web}, where, in terms of service interaction, routing patterns and messaging between Web applications become feasible.
A cloud PaaS that relies on webhooks for distributed asynchronous processing is Google's App Engine \cite{google-pushqueues}.

Another approach to syndication is by so-called feeds or channels.
Clients and Web applications can subscribe to a feed offered by a syndication service to receive updates, i.e., a service-side One-to-Many Send interaction pattern.
Two XML-based standards for feeds are the Rich Site Summary (RSS) \cite{rss} and the Atom syndication format~\cite{RFC4287}.
While RSS assumes HTTP as transport mechanism, Atom defines its own HTTP-based publishing protocol (AtomPub)~\cite{RFC5023}.

A syndication service needs to initiate communication with the client when updates are available.
A Web client can either poll the syndication service's feed or use HTTP push technology.
For service-to-service interaction, a Web application registers a webhook at the syndication service to receive updates when available.
A syndication service that pushes updates to clients has the role of a broker in a publish-subscribe architecture as shown in Fig. \ref{fig:pubsub}.
Two implementations for syndication are PubSubHubBub~\cite{pubsubhubbub} and the AtomPub-oriented Apache Abdera~\cite{apache-abdera}.

The Bayeux protocol \cite{bayeux} is a push framework based on Comet using named channels for a broker-based publish-subscribe architecture, i.e., One-to-Many Send interaction.
Clients subscribe to named channels, and the server pushes updates to all registered clients.
An application of Bayeux is Web feeds.

\subsubsection{Web mashup}
\label{sub:mashup}

A mashup composes so-called Web components, e.g., multimedia resources or script code, from different origins into a new website \cite{Endres-Niggemeyer2013}.
Examples for mashup components are JavaScript libraries, gadgets \cite{google-gadgets}, and services like Google Maps.
Web components and the mashup principle achieve composability and reusability of resources which is also a contributing factor for the success of the Web 2.0.
A mashup service, also called integrator, can be distinguished based on the location, where integration of Web components takes place \cite{Ryck2012,Soylu2012}:

\begin{itemize}
	\item \emph{Service-side mashup.} \ When a client requests a mashed-up service, the service first gathers the foreign resources from other origins, processes them, and returns the integrated markup and media to the client.
	\item \emph{Client-side mashup.} \ A client-side mashup service returns a markup skeleton and script code, so the client embeds components statically or loads them dynamically using AJAX.
	The client is then responsible for integration.
\end{itemize}

\begin{figure}
	\begin{center}
	\subfigure[Client-side mashup.]{
\begin{tikzpicture}[->, >=latex, box/.style={draw,align=center, minimum height=5.5mm}]
\node[box] (v1) at (0.4,1.2) {\textsf{Mashup}};
\node[box] (v3) at (1.2,2.3) {$C_1$};
\node[box] (v5) at (2.4,2.3) {$C_n$};
\node at (1.8,2.3) {$\dots$};
\draw  (0.1,0.4) rectangle (2.5,-0.3);
\draw[dotted]  (2,0.3) rectangle (2.4,-0.2);
\draw[dotted]  (1.2,0.3) rectangle (1.6,-0.2);
\node at (1.8,0) {$\dots$};
\node at (0.6,0) {\textsf{Client}};
\node (v4) at (1.4,0.3) {};
\node (v6) at (2.2,0.3) {};
\node (v2) at (0.4,0.3) {};
\draw  (v1) edge (v2);
\draw  (v3) edge (v4);
\draw  (v5) edge (v6);
\end{tikzpicture}
		\label{fig:client_mashup}
	}
	\hspace{3em}
	\subfigure[Service-side mashup.]{
\begin{tikzpicture}[->, >=latex, box/.style={draw,align=center, minimum height=5.5mm}]
\node[box] (v3) at (0,2.3) {$C_1$};
\node[box] (v5) at (1.2,2.3) {$C_n$};
\node at (0.6,2.3) {$\dots$};
\draw  (-1.3,1.5) rectangle (1.3,0.8);
\draw[dotted]  (0.8,1.4) rectangle (1.2,0.9);
\draw[dotted]  (0,1.4) rectangle (0.4,0.9);
\node at (0.6,1.1) {$\dots$};
\node at (-0.7,1.1) {\textsf{Mashup}};
\node (v4) at (0.2,1.4) {};
\node (v6) at (1,1.4) {};
\draw  (v3) edge (v4);
\draw  (v5) edge (v6);
\node[box] (v2) at (0,0) {\textsf{Client}};
\node (v1) at (0,0.9) {};
\draw  (v1) edge (v2);
\end{tikzpicture}
		\label{fig:service_mashup}
		\hspace{0.1em}
	}
	\end{center}
\vspace{-1\baselineskip}
\caption{A mashup integrates $n$ Web resources from different origins as Web components $C_1 \dots C_n$ \cite{Ryck2012}.}
\label{fig:mashup}
\end{figure}
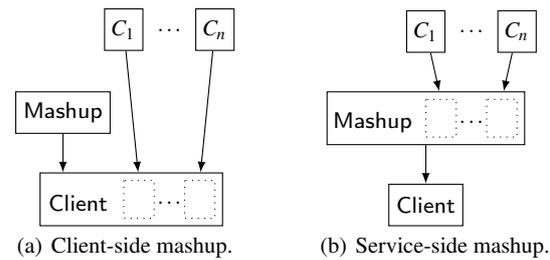

Fig. \ref{fig:mashup} shows both cases.
Script code acts as a glue between Web components, and the information flow between components can lead to security issues.
Web components therefore require proper encapsulation \cite{Magazinius2013}.
In terms of patterns, a mashup enables one or more parallel bilateral Send-Receive interactions between a client and Web servers that host Web components, eventually routing messages between them.

Client-side content policies lead to two extreme cases of local script code interaction: no separation nor isolation by direct embedding in the same DOM using a \texttt{script} tag and strong separation and isolation by embedding in isolated DOMs using \texttt{object} or \texttt{iframe} elements.
Ryck et al. \cite{Ryck2012} survey the state-of-the-art techniques for more fine-grained controls in mashups, and they distinguish four categories of Web component integration restrictions:

\begin{enumerate}
	\item \emph{Separation and interaction.} \ Components are separated such that individual component DOMs and script code, e.g., global variables, are fully isolated from each other.
	Components can then interact through specified channels, e.g., HTML5 offers the \texttt{sandbox} attribute for fine-grained \texttt{iframe} separation and Web Messaging for interaction between \texttt{iframe} objects.
	 \item \emph{Script isolation.} \ To isolate script code in components when running in the same runtime environment, code can be restricted to a subset of allowed functions, and static checking enforces an isolation policy.
	 \item \emph{Cross-domain communication.} \ The same-origin policy denies cross-domain communication for script code by default during runtime.
	 Techniques like CSP, CORS, or an \texttt{XMLHttpRequest} proxy allow read access according to a defined policy.
	 \item \emph{Behavior control.} \ This category subsumes techniques to enforce policies on script code execution during runtime, e.g., reference monitors, access mediation to objects, or information flow control.
\end{enumerate}

To ease composability of mashup components, public API specifications, also referred to as \emph{Open APIs}, have become popular in recent years.
Open APIs can range from Web resource access to sophisticated service architectures as discussed in Sect. \ref{sub:serviceview}.
In particular, OpenSocial \cite{opensocial} is an initiative to standardize APIs for building social Web applications.
Integrating the social dimension is a step toward personalized user experience, which is a characteristic of the Web 3.0.

\subsection{Service-oriented view}
\label{sub:serviceview}
Custom network protocols for enterprise application integration are often not forwarded over the Internet because middleboxes filter them.
Using Web technology for integration, e.g., in a middleware approach, ensures service availability across the Internet.
In our definition, a service has an interface that accepts and responds with a certain language.
According to the interface, services can be distinguished by:

\begin{itemize}
	\item \emph{Static typing.} \ When the accepted language of a service is predefined by an Interface Definition Language (IDL) or a grammar (e.g., a specific MIME media type, a schema in XML, or a specification in ASN.1), the interface is strict.
	An implementation can be automatically generated for the accepted language, e.g., a parser or stub code.
	\item \emph{Dynamic typing.} \ On the other hand, a service eventually accepts a set of content types, a language family, or messages carry their own specifications, e.g., embedded schemas.
	Interpretation is then runtime dependent, and parsers need to be modular.
	Such a service interface is referred to as dynamically typed.
\end{itemize}

Typing affects coupling between clients and services.
Web applications, syndication, and mashups in the previous section are dynamically typed because the content type of a resource governs the selection of a proper parser and semantics in the client.
Web applications are loosely coupled.

Statically typed service interfaces restrict flexibility and interoperability, e.g., the allowed programming languages, where stub code is available.
The consequence is tighter coupling.
Services and middleware can still benefit from loose coupling because evolving software systems become easier and cheaper to integrate \cite{Pautasso2009a}.
This section surveys architectures for offering services in cloud environments.
Four architectures are discussed: RPC, ``big'' Web services, RESTful Web services, and message-oriented middleware.

\subsubsection{Remote procedure calls}
\label{sub:rpc}

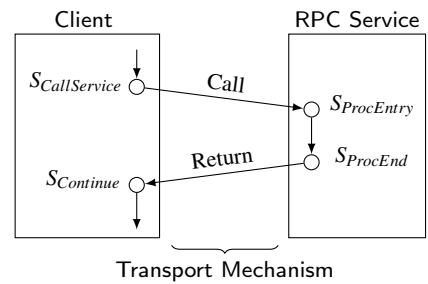
\begin{figure}
	\begin{center}
\usetikzlibrary{decorations.pathreplacing}
\begin{tikzpicture}[anchor=base,->, >=latex]
\tikzstyle{res}=[draw, dotted, anchor=west, minimum height=4.5mm]
\tikzstyle{nod}=[circle, draw, inner sep=2pt, label distance=-4pt]

\draw  (1.6,2.8) rectangle (3.4,0.1);
\node at (2.5,2.9) {\textsf{RPC Service}};
\draw  (-2,2.8) rectangle (-0.1,0.1);
\node at (-1.1,2.9) {\textsf{Client}};
\node[nod] (v2) at (-0.4,2.1) {};

\node at (-1.2,2.1) {$S_{CallService}$};

\node[nod] (v4) at (-0.4,0.8) {};

\node at (-1.1,0.8) {$S_{Continue}$};

\draw [-,decorate,decoration={brace,amplitude=3pt},xshift=-4pt,yshift=0pt]
(1.6,0) -- (0.2,0) node [black,below,midway,yshift=-1mm, align=center] {\textsf{Transport Mechanism}};
\node[nod] (v5) at (1.9,1.8) {};
\node[nod] (v6) at (1.9,1.1) {};
\node at (2.7,1.8) {$S_{ProcEntry}$};
\node (v1) at (-0.4,2.7) {};
\node (v3) at (-0.4,0.1) {};
\draw (v1) edge (v2);
\draw (v4) edge (v3);
\node at (2.7,1.1) {$S_{ProcEnd}$};
\draw  (v5) edge (v6);
\draw  (v2) edge node [sloped, above] {Call} (v5);
\draw  (v6) edge node [sloped, above] {Return} (v4);
\end{tikzpicture}
	\end{center}
\caption{RPC is conceptually a Send-Receive interaction. A client serializes the arguments for a remote function call as a message, sends it to the RPC service, and waits for a response.}
\label{fig:rpc}
\end{figure}

RPC is a simple yet powerful architectural style to offer a service by exposing network-accessible functions \cite{Birrell1984}.
In terms of patterns, RPC is bilateral Send-Receive between a client and a service, as shown in Fig. \ref{fig:rpc}. 
The client initiates the interaction, and if not stated otherwise, a network function call in RPC is synchronous and interfaces are statically typed.
Specifying an RPC service requires an agreed-upon transport mechanism, e.g., TCP or HTTP, an agreement on how to address and bind to a remote function, and a data serialization format to exchange structured data, e.g., ASN.1. 

Historically, one of the most widely deployed RPC solutions is Open Network Computing RPC (ONC-RPC) \cite{RFC5531}, e.g., for network file systems.
ONC-RPC originates from Sun Microsystems, and APIs are available on practically all major platforms.
ONC-RPC uses TCP and UDP as transport mechanism, where call and return values are serialized in the XDR format.

RPC for distributed systems has evolved from function calls to distributed computation over shared objects~\cite{Vogels2003}.
Zarras~\cite{Zarras2004} analyzes three prominent RPC-style middleware approaches that are based on object sharing:

\begin{itemize}
	\item Microsoft Component Services (COM+) \cite{ms-complus};
	\item The OMG-standardized Common Object Request Broker Architecture (CORBA)~\cite{omg-corba}; and
	\item Java Remote Method Invocation (RMI) \cite{oracle-jrmi} in the Java Platform, Enterprise Edition (Java EE) \cite{oracle-javaee}.
\end{itemize}

All three approaches define individual data serialization formats and support communication over the Internet protocols, in particular, TCP.
While protocols and wire formats for COM+ are specified in the DCE/RPC standard~\cite{dcerpc}, CORBA uses the Internet Inter-ORB Protocol (IIOP)~\cite{omg-iiop} for communication over TCP connections.
Java RMI specifies individual protocols and wire formats on top of TCP, e.g., the Java Remote Method Protocol (JRMP) or Oracle Remote Method Invocation (ORMI), but also supports RMI over IIOP~\cite{oracle-rmi-iiop} for compatibility with CORBA systems.

The three approaches enable shared objects in an RPC-style architecture.
An object broker for allocation, garbage collection, and transactions is implicitly required \cite{Vogels2003}.
To represent a shared object on the client side, a stub abstracts away the serialization and communication.
The service interface is therefore statically typed.
A recent middleware framework, similar to CORBA, but compatible with Web protocols, is the Internet Communications Engine \cite{zeroc-ice}.

\paragraph{XML- and JSON-RPC.}  A basic architecture for a Web-based RPC is XML-RPC \cite{xmlrpc}.
Call and return information is serialized as text-based XML, and HTTP transports request and response documents to a service identified by a URL.
While XML-RPC defines a subset of XML by specifying rules how datatypes are serialized using tags and attributes, an XML-RPC service is still dynamically typed because there is no schema for an actual service interface.
The XML-RPC Description Language (XRDL) \cite{xrdl} is an attempt to define XML-RPC services in the spirit of an IDL, so client-side stub code can be automatically generated.

JSON-RPC \cite{jsonrpc} is related to XML-RPC, with the difference that the specification does not restrict itself to any transport mechanism.
JSON-RPC only specifies call and return formats based on the JSON format.

\paragraph{Apache Avro, Etch, and Thrift.} \ Several proprietary implementations for RPC-style service interaction have been proposed, especially for high-performance provider-side backend services in clouds.
Apache Avro \cite{apache-avro} is both a data serialization format and an RPC mechanism in the Apache Hadoop \cite{apache-hadoop} project for big data analysis.
Transportation of messages in Avro is protocol independent in general, but the specification explicitly refers to HTTP.
Avro supports dynamic typing; a serialized message is accompanied by its schema in JSON format.

The Apache Etch \cite{apache-etch} framework originates from Cisco Systems.
It defines a service description language, similar to an IDL, a binary serialization format, and TCP is assumed for transportation.
A compiler generates stub code for the client and service sides of a specified architecture, and messages are therefore statically typed.

Apache Thrift \cite{apache-thrift}, developed by Facebook, is a framework that specifies an IDL for RPC-style services and tools for statically typed stub code.
While transportation and serialization of messages is kept abstract, Thrift provides several default procedures, e.g., serialization in a compact binary format or JSON and transport over TCP or HTTP.
A notable RPC framework based on Thrift is Twitter Finagle \cite{twitter-finagle}. 
Finagle enhances Thrift with multiplexing; Twitter's Mux session-layer protocol is situated between TCP and Thrift, so only a single connection between the client and the service is necessary to transport parallel Thrift interactions.

\paragraph{Protocol Buffers.} \ Google Protocol Buffers \cite{google-protobuf} define a platform-neutral language for specifying a data structure, similar to ASN.1.
Tools translate a specified structure into stub code for compact binary serialization and deserialization.
RPC services using Protocol Buffers are therefore statically typed.
There are no restrictions on transport mechanisms or message exchange.
As long as both parties have stub code generated for the same specified structure, messages can be exchanged, e.g., over HTTP.
Google claims that their internal RPC services rely on Protocol Buffers~\cite{google-protobuf}.

\paragraph{Hessian and Burlap.} \ Hessian \cite{caucho-hessian}  is a protocol for RPC Web-based services, where structured information is serialized in a compact binary format.
This message format is dynamically typed, designed for efficient processing, and intended for HTTP transportation.
No IDL is therefore necessary for a Hessian service.
Burlap \cite{caucho-burlap} is semantically the same protocol as Hessian, but uses an XML-based data serialization format.

\subsubsection{SOAP/WS-* Web services}
\label{sub:WebServices}

Many RPC architectures require static typing in an IDL and, as a consequence, have tight coupling from code restrictions.
The goal of so-called big Web services is to relax this coupling by open standards for heterogeneous platforms \cite{Pautasso2009a}.
A Web service deals with XML documents and document encapsulation to evade the complexity of distributed computation on shared objects \cite{Vogels2003}.
The core technology in this attempt is the Simple Object Access Protocol (SOAP)~\cite{w3c-soap} for expressing messages as XML documents.

SOAP is transport-agnostic by design and supports all kinds of transport mechanisms, including message-oriented middleware.
However, HTTP has become an industry standard for because of its middlebox compatibility~\cite{Gottschalk2002}.
Web services standards (referred to as WS-*) extend SOAP-based interaction with security, reliability, transaction, orchestration, workflow, and business process aspects.

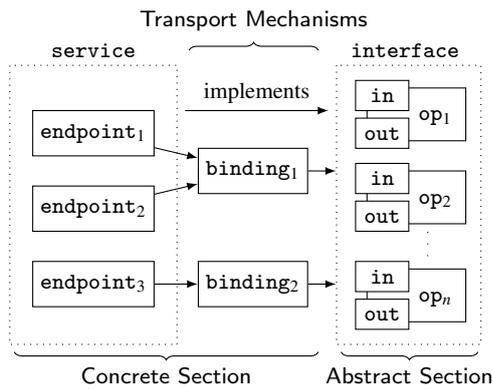
\begin{figure}
	\begin{center}
\usetikzlibrary{decorations.pathreplacing}
\begin{tikzpicture}[->, >=latex]
\tikzstyle{io}=[draw,fill=white, minimum width=7mm, minimum height=4mm]
\tikzstyle{bind}=[draw, minimum width=8mm, minimum height=6mm]
\tikzstyle{end}=[draw, minimum width=8mm, minimum height=6mm]

\node at (2.5,3.1) {\texttt{op}$_1$};
\node at (2.5,2) {\texttt{op}$_2$};
\node at (2.5,0.7) {\texttt{op}$_n$};
\node (v8) at (2.4,1.7) {};
\node (v9) at (2.4,1.1) {};
\draw[-, loosely dotted ]  (v8) edge (v9);

\draw  (1.6,3.5) rectangle (2.9,2.8);
\draw  (1.6,2.4) rectangle (2.9,1.7);
\draw  (1.6,1.1) rectangle (2.9,0.4);
\node[io] at (1.8,3.4) {\texttt{in}};
\node[io] at (1.8,2.9) {\texttt{out}};
\node[io] at (1.8,2.3) {\texttt{in}};
\node[io] at (1.8,1.8) {\texttt{out}};
\node[io] at (1.8,1) {\texttt{in}};
\node[io] at (1.8,0.5) {\texttt{out}};
\draw[dotted]  (1.2,3.8) rectangle (3.1,0.1);
\node (v26) at (2.1,4) {\texttt{interface}};
\node[bind] (v18) at (0.1,2.4) {\texttt{binding}$_1$};
\node[bind] (v20) at (0.1,0.9) {\texttt{binding}$_2$};

\node[end] (v17) at (-2,2.9) {\texttt{endpoint}$_1$};
\node[end] (v19) at (-2,1.9) {\texttt{endpoint}$_2$};
\node[end] (v21) at (-2,0.9) {\texttt{endpoint}$_3$};

\node (v25) at (-2,4) {\texttt{service}};

\draw[dotted]  (-0.9,0.1) rectangle (-3.1,3.8);

\draw [-,decorate,decoration={brace,amplitude=3pt},xshift=-4pt,yshift=0pt]
(3.2,0) -- (1.4,0) node [black,below,midway,yshift=-1mm, align=center] {\textsf{Abstract Section}};

\draw [-,decorate,decoration={brace,amplitude=3pt},xshift=-4pt,yshift=0pt]
(1.1,0) -- (-2.9,0) node [black,below,midway,yshift=-1mm, align=center] {\textsf{Concrete Section}};

\draw [-,decorate,decoration={brace,amplitude=3pt},xshift=-4pt,yshift=0pt]
(-0.6,4) -- (1.1,4) node [black,above,midway,yshift=1.5mm, align=center] {\textsf{Transport Mechanisms}};

\node (v1) at (1.3,2.4) {};
\node (v2) at (1.3,0.9) {};
\node (v3) at (-0.9,3.2) {};
\node (v4) at (1.2,3.2) {};
\draw  (v3) edge [] node[above] {implements} (v4);
\draw  (v17) edge (v18);
\draw  (v18) edge (v1);
\draw  (v19) edge (v18);
\draw  (v21) edge (v20);
\draw  (v20) edge (v2);
\end{tikzpicture}
	\end{center}
\caption{A WSDL 2.0 service definition characterizes an interface as a set of operations. A binding assigns a concrete transport mechanism to an abstract interface, e.g., SOAP over HTTP. A service then implements an interface in a set of endpoints, i.e., addresses, for bindings.}
\label{fig:wsinterface}
\end{figure}

\begin{figure}
	\begin{center}
\usetikzlibrary{decorations.pathreplacing}
\begin{tikzpicture}[anchor=base,>=latex]

\node (v4) at (6.9,3.9) {Coordination};
\node (v5) at (6.9,4.6) {Security, Reliability, Transactions};
\node (v5) at (6.9,5.1) {Composition, Choreography, Orchestration};

\draw[dotted]  (4.3,3.7) rectangle (9.5,4.3);
\draw[dotted]  (9.5,5.5) rectangle (4.3,4.4);

\node at (6.9,0.8) {TCP};
\node at (6.9,1.4) {SSL/TLS};
\node at (6.9,2) {HTTP};
\node at (6.9,2.6) {SOAP Msg. Security};
\node at (6.9,3.2) {SOAP};

\draw  (4.3,2.4) rectangle (9.5,1.8);
\draw   (9.5,1.8) rectangle (4.3,1.2);
\draw  (4.3,1.2)  rectangle (9.5,0.6);
\draw  (4.3,3) rectangle (9.5,2.4);
\draw  (4.3,3.6) rectangle (9.5,3);

\draw [decorate,decoration={brace,amplitude=3pt},xshift=-4pt,yshift=0pt]
(4.3,2.5) -- (4.3,3.6) node [black,left,midway,xshift=-1mm, align=center] {\textsf{Messaging}};

\draw [decorate,decoration={brace,amplitude=3pt},xshift=-4pt,yshift=0pt]
(4.3,0.6) -- (4.3,2.4) node [black,left,midway,xshift=-1mm, align=center] {\textsf{Transport}};
\draw [decorate,decoration={brace,amplitude=3pt},xshift=-4pt,yshift=0pt]
(4.3,3.7) -- (4.3,4.3) node [black,left,midway,xshift=-1mm, align=center] {\textsf{Protocol}\\\textsf{Infrastructure}};
\draw [decorate,decoration={brace,amplitude=3pt},xshift=-4pt,yshift=0pt]
(4.3,4.4) -- (4.3,5.5) node [black,left,midway,xshift=-1mm, align=center] {\textsf{Middleware}\\\textsf{Properties}};
\draw[dotted]  (9.5,0.5) rectangle (4.3,-0.1);
\node at (7,0.1) {Internet/Networking};

\end{tikzpicture}
	\end{center}
\caption{The Web services interaction stack \cite{Alonso2004} in accordance with the WS-I Basic Profile \cite{wsi-profile12,wsi-profile20} is restricted to SOAP over HTTP as transport mechanism}
\label{fig:webservices}
\end{figure}
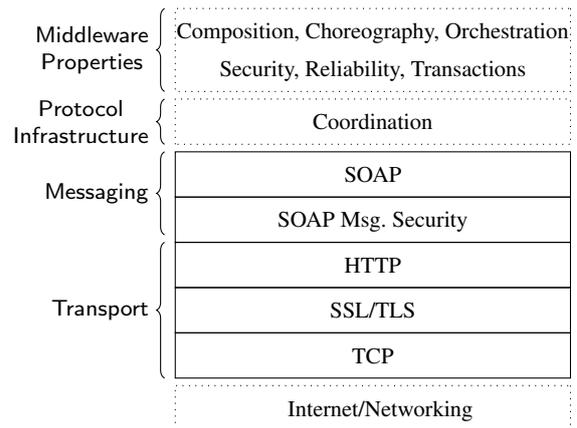

\paragraph{Technologies.} \ In accordance with Alonso et al. \cite{Alonso2004}, technologies for Web services are categorized into four groups:

\begin{itemize}
	\item \emph{Service description.} \ The XML-based Web Services Description Language (WSDL) \cite{w3c-wsdl} is a format to describe Web services similar to an IDL.
	While coupling between Web services is loose due to XML, interfaces and messages in the SOAP/WS-* stack are nonetheless statically typed in WSDL.
	
	A WSDL version 2.0 document has an abstract and a concrete section, as shown in Fig. \ref{fig:wsinterface}.
	The abstract section defines a \texttt{type} system of one or more XSDs to specify message formats; one or more \texttt{interfaces} and their \texttt{operations}; and an assignment of \texttt{input} and \texttt{output} message types to operations.
	Every operation has a message exchange \texttt{pattern}: \texttt{in} for Receive, \texttt{out} for Send, and \texttt{in-out} for Receive-Send interaction.
	
	The concrete section in WSDL describes how abstract interfaces become network accessible.
	A \texttt{binding} associates a specific transportation mechanism, e.g., SOAP over HTTP, to an abstract interface.
	Finally, a \texttt{service} implements an abstract interface by specifying a set of endpoints.
	An \texttt{endpoint} associates an accessible address (URL), where the service can be consumed, to a concrete binding.
	
	\item \emph{Service discovery.} \ Services described in WSDL need to be published to enable automatic discovery or dynamic late binding.
	The XML-based Universal Description, Discovery, and Integration (UDDI) \cite{oasis-uddi} standardizes a registry for classifying, cataloging, and managing Web services.
	Such a registry is typically offered as a SOAP/WS-* Web service.

	\item \emph{Service interaction.} \ To access operations, clients and services need to communicate SOAP messages.
	Alonso et al. \cite{Alonso2004} introduce the notion of a service interaction stack, as shown in Fig. \ref{fig:webservices}.
	The stack has four layers: \emph{transport} of messages by HTTP, eventually protected by SSL/TLS; \emph{messaging} through SOAP; a \emph{protocol infrastructure} to coordinate a number of services using meta-protocols; and \emph{middleware properties} with respect to security, reliability, transactions, and orchestration of services.
	
	\item \emph{Service composition.} \ Network-accessible operations allow Web services to be composed.
	In terms of interaction patterns, a basic Web service that computes a result implements Send, Receive, or Receive-Send.
	A composite service consumes other services, and according to the application logic, it allows more advanced interaction patterns.
\end{itemize}

Due to the large number and many versions of WS-* standards, the Web Services Interoperability Organization (WS-I) establishes best practices for interoperability, published as Basic Profiles \cite{wsi-profile12,wsi-profile20}.
Figure \ref{fig:webservices} is limited to basic profile protocols and standards for transport and messaging.

\paragraph{SOAP attachments.} \ A SOAP message has an \texttt{envelope} root element that holds a \texttt{header} element for metadata and a \texttt{body} element for the message content.
The structure of the body has to obey the schema of its according WSDL message type, and a service can therefore validate messages.

Base64 allows to encapsulate arbitrary binary data as XML-compatible text, but incurs a blowup in size.
Relevant metadata, e.g., the MIME type of the binary content, is not preserved in a standardized way.
Several techniques have therefore been proposed for dealing with binary content in SOAP.

Two historical standards are SOAP Messages with Attachments (SwA) \cite{w3c-soap-attachment-old,w3c-soap-attachment} and Microsoft's DIME.
SwA relies on MIME multipart \cite{RFC2387} as container format, where XML-based SOAP is the first part, and subsequent parts have individual MIME types and encapsulate binary contents directly.
DIME operates in the same spirit but with a different container format.
As there are different types of MIME multipart containers for SwA, the WS-I has published the Attachments Profile \cite{wsi-attachments} for interoperability.

The state-of-the-art SOAP Message Transmission Optimization Mechanism (MTOM)~\cite{w3c-mtom} resorts to XOP packaging if a WSDL input message type defines an element annotated with a certain MIME content type.
From the service's WSDL, the client becomes aware that MTOM is accepted.
If a client does not support MTOM, Based64 is used as fallback encoding.

\paragraph{Messaging.} \ The SOAP standard does not specify how messages are routed, whom to respond to in asynchronous communication, or where to report errors.
SOAP is in a sense similar to RPC; the operation is completely defined by its endpoint address, e.g., a URL for a HTTP transport binding.

WS-Addressing \cite{w3c-wsaddressing} specifies two core constructs to enable routing patterns: endpoint references and message addressing.
An endpoint reference is an address, i.e., URL, and optional parameters for communicating with the endpoint.
Addressing information is stored in the SOAP header, e.g., a semantic message identifier (\texttt{wsa:Action}), destination endpoint references (\texttt{wsa:To}), routing or relay endpoint references (\texttt{wsa:ReplyTo}), or endpoints for error handling (\texttt{wsa:FaultTo}).
Based on addressing information, a service can dynamically forward SOAP messages to other services to implement all kinds of interaction patterns.

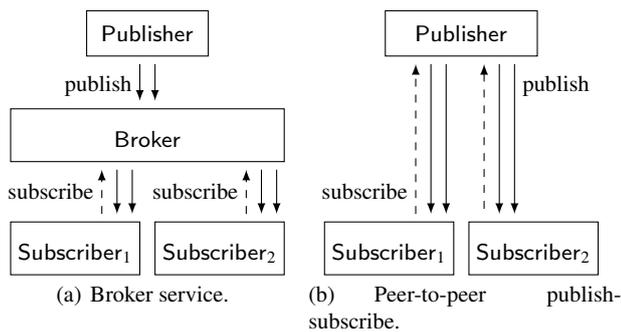
\begin{figure}
	\begin{center}
	\subfigure[Broker service.]{
\begin{tikzpicture}[anchor=base,>=latex]
\node (v1) at (3.2,3.7) {\textsf{Publisher}};
\node (v4) at (2.25,0.8) {\textsf{Subscriber}$_1$};
\node (v2) at (3.2,2.3) {\textsf{Broker}};
\node (v3) at (4.15,0.8) {\textsf{Subscriber}$_2$};
\draw  (1.4,2.8) rectangle (5,2.1);

\node (v9) at (3.1,2.8) {};
\node (v12) at (4.7,2.1) {};
\node (v10) at (2.8,2.1) {};
\node (v6) at (2.6,2.1) {};
\node (v8) at (4.5,2.1) {};
\node (v5) at (2.6,1.3) {};
\node (v11) at (2.8,1.3) {};
\node (v7) at (4.5,1.3) {};
\node (v13) at (4.7,1.3) {};
\draw[dashed,->]  (v5) edge node[left] {subscribe} (v6);
\draw[dashed,->]  (v7) edge node[left] {subscribe} (v8);

\draw[->]  (v10) edge (v11);
\draw[->]  (v12) edge (v13);
\draw  (2.4,4.1) rectangle (4,3.5);
\node (v14) at (3.1,3.5) {};
\draw[->]  (v14) edge node[left,yshift=-0.5mm] {publish} (v9);
\draw  (1.4,1.3) rectangle (3.1,0.6);
\draw  (3.3,1.3) rectangle (5,0.6);
\node (v15) at (3.3,3.5) {};
\node (v16) at (3.3,2.8) {};
\draw[->]  (v15) edge (v16);
\node (v17) at (3,2.1) {};
\node (v18) at (3,1.3) {};
\node (v19) at (4.9,2.1) {};
\node (v20) at (4.9,1.3) {};
\draw[->]  (v17) edge (v18);
\draw[->]  (v19) edge (v20);
\end{tikzpicture}
		\label{fig:pubsub_broker}
	}
	\hspace{0.3em}
	\subfigure[Peer-to-peer publish-subscribe.]{
\begin{tikzpicture}[anchor=base,>=latex]
\node (v1) at (3.2,3.7) {\textsf{Publisher}};
\node (v4) at (2.25,0.8) {\textsf{Subscriber}$_1$};
\node (v3) at (4.15,0.8) {\textsf{Subscriber}$_2$};
\node (v9) at (3.9,1.3) {};
\node (v6) at (2.6,3.5) {};
\node (v5) at (2.6,1.3) {};

\draw[dashed,->]  (v5) edge node[left,yshift=-7mm] {subscribe} (v6);

\draw  (2.2,4.1) rectangle (4.2,3.5);
\node (v14) at (3.9,3.5) {};
\draw[->]  (v14) edge node[right,yshift=7mm] {publish} (v9);
\draw  (1.4,1.3) rectangle (3.1,0.6);
\draw  (3.3,1.3) rectangle (5,0.6);
\node (v15) at (2.8,3.5) {};
\node (v16) at (2.8,1.3) {};
\draw[->]  (v15) edge (v16);

\node (v11) at (3.5,3.5) {};

\node (v12) at (3.5,1.3) {};

\draw[<-, dashed]  (v11) edge (v12);

\draw (3,3.5) node (v2) {};
\node (v7) at (3,1.3) {};
\draw[->]  (v2) edge (v7);

\node (v8) at (3.7,3.5) {};
\node (v10) at (3.7,1.3) {};
\draw[->]  (v8) edge (v10);
\end{tikzpicture}
		\label{fig:pubsub_p2p}
		\hspace{0.5em}
	}
	\end{center}
\vspace{-1\baselineskip}
\caption{Publish-subscribe can either use a broker- or peer-to-peer-based architecture to facilitate multilateral service interaction patterns}
\label{fig:pubsub}
\end{figure}

WS-Eventing \cite{w3c-eventing} and WS-Notification \cite{oasis-wsnotification} are two competing standards for publish-subscribe messaging on top of WS-Addressing.
In terms of interaction patterns, publish-subscribe is One-to-Many Send from point of view of a publisher.
Publish-subscribe either needs a broker service or the publisher manages subscriptions individually, as shown in Fig. \ref{fig:pubsub}.
A broker stores subscribed endpoint references and distributes messages from publishers.
So, broker-based pub\-lish-subscribe allows a publisher to address an anonymous group of receivers.

\paragraph{Security and reliability.} \ SSL/TLS in HTTPS bindings in SOAP only ensures encryption between the client and service endpoint, but according to WS-Addressing, a SOAP message could traverse multiple services or brokers until its destination is met.
WS-Security~\cite{oasis-wssecurity} defines standards for end-to-end cryptography methods, including XML signatures and encryption for SOAP messages.
WS-Trust~\cite{oasis-wstrust} and WS-SecureConversation~\cite{oasis-wssecureconversation} extend WS-Security by establishing a security context for communicating parties to speed up the cryptographic key exchange.

The WS-Policy \cite{w3c-wspolicy} framework for Web services defines a language for specifying and advertising policy assertions, e.g., security or Quality-of-Service policies.
WS-SecurityPolicy \cite{oasis-wssecuritypolicy} is an extension of WS-Security to express security-specific policies.

Another issue that arises in a messaging context is reliable message delivery.
WS-Reliability and its successor WS-ReliableMessaging \cite{oasis-wsreliability} specify procedures, so delivery is guaranteed through acknowledgments even for routed SOAP messages.

\paragraph{Protocol infrastructure and middleware properties.} \ To exploit SOAP messaging in SOA, standards for coordinating and composing services to model business processes are required.
A notable standard for such a protocol infrastructure is WS-Coordination \cite{oasis-wscoordination} to coordinate actions of multiple Web services.
WS-Coordination enables distributed activities like transactions (WS-AtomicTransaction \cite{oasis-wstransaction}) or business activities (WS-BusinessActivity \cite{oasis-wsba}).

Beraka et al.~\cite{Beraka2012} review standards for Web service composition and orchestration.
Examples for composition and orchestration modeling are the Business Process Model and Notation (BPMN)~\cite{omg-bpmn}, Business Process Execution Language (BPEL)~\cite{oasis-bpel} and WS-Choreography~\cite{w3c-wschoreography}.

\paragraph{Software implementations.} \ In the Java world, there are several APIs supporting SOAP/WS-* Web services integration like JAXM \cite{java-jaxm} for SOAP integration, the obsoleted JAX-RPC \cite{java-jax-rpc}, and its successor JAX-WS \cite{java-jax-ws} for Web services.
The most notable implementations using those APIs are Apache Axis2 \cite{apache-axis}, Apache CXF \cite{apache-cxf}, GlassFish \cite{java-metro}, IBM WebSphere \cite{ibm-jaxws}, JBoss \cite{jbossws}, Oracle Weblogic \cite{oracle-weblogic}, and XML Interface for Network Services (XINS) \cite{xins}.

Furthermore, the Windows Communication Framework (WCF) \cite{ms-wcf} is an API and runtime environment in Microsoft's .NET framework that relies on SOAP/WS-* for services integration.

\subsubsection{RESTful services}
\label{sub:RestfulWS}

Fielding \cite{Fielding2000} has coined the term ``Representational State Transfer'' (REST) as an architectural style for distributed hypermedia systems.
REST for resource-oriented architectures has become a key technique in the Web and cloud services to achieve simplicity, scalability, and shareability.
In its original form, REST is rather abstract and not restricted to specific protocols.
However, the principles have been derived from successful Web architectures and early versions of HTTP.
That is why REST primarily refers to HTTP as transport mechanism today.

\begin{figure}
	\begin{center}
\usetikzlibrary{decorations.pathreplacing}
\begin{tikzpicture}[anchor=base,->, >=latex]
\tikzstyle{res}=[draw, dotted, anchor=west, minimum height=4.5mm]
\tikzstyle{nod}=[circle, draw, inner sep=2pt, label distance=-4pt]

\node[res] (v1) at (1.9,2.6) {\texttt{/cart}};
\node[res] (v5) at (1.9,2) {\texttt{/product/1}};
\node[res] at (1.9,1.4) {\texttt{/product/2}};
\node[res] (v6) at (1.9,0.8) {\texttt{/product/3}};
\node[res] (v7) at (1.9,0.2) {\texttt{/order}};
\draw  (1.6,3) rectangle (4,-0.2);
\node at (2.8,3.1) {\textsf{RESTful Service}};
\draw  (-2.6,2.7) rectangle (-0.4,0.2);
\node at (-1.6,2.8) {\textsf{Client}};
\node[nod] (v2) at (-0.8,2.3) {};
\draw  (v1) edge node [above] {\texttt{GET}} (v2);
\node at (-1.5,2.3) {$S_{NeedCart}$};
\node[nod] (v3) at (-0.8,1.4) {};
\node[nod] (v4) at (-0.8,0.6) {};
\draw  (v2) edge (v3);
\draw  (v3) edge (v4);
\draw (v3) edge[loop left, looseness=40] (v3);
\node at (-1.7,1.7) {$S_{AddProducts}$};
\node at (-1.4,0.6) {$S_{Order}$};
\draw  (v5) edge node [above] {\texttt{GET}} (v3);
\draw  (v6) edge node [above] {\texttt{GET}} (v3);
\draw  (v4) edge node [above] {\texttt{PUT}} (v7);
\draw [-,decorate,decoration={brace,amplitude=3pt},xshift=-4pt,yshift=0pt]
(1.6,-0.3) -- (-0.1,-0.3) node [black,below,midway,yshift=-1mm, align=center] {\textsf{HTTP Interaction}};
\end{tikzpicture}
	\end{center}
\caption{A client interacts with a RESTful service, keeps track of application state locally, and applies HTTP methods on resources. From the \texttt{cart}, the client discovers URLs of products and order placements.}
\label{fig:rest}
\end{figure}
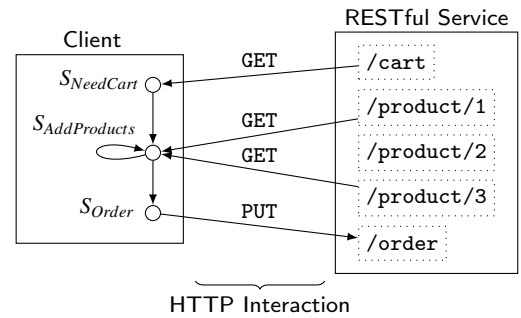

REST emphasizes a \emph{unified interface} between components and abstraction of information as resource.
Four interface constraints for an architectural style to be considered RESTful are established by Fielding~\cite{Fielding2000,Fielding2002}:

\begin{enumerate}
	\item \emph{Identification of resources.} \ A REST service offers a set of resources that need to be identified, so clients can interact with them.
	URIs are well known for resource identification in the Web and also for REST. HTTP is the protocol of choice for interaction.
	
	\item \emph{Manipulation of resources through representations.}
	
	A representation of a resource is a sequence of bytes plus metadata that describe those bytes using a content type.
	A representation captures the current state of a resource, and REST operations applied to a representation modify the resource's state.
	REST reuses HTTP methods as unified operations on resources.
	So, existing infrastructure such as caches or proxies stay compatible with REST, which in turn gives high scalability for PaaS or SaaS environments.
	The following HTTP methods are defined as operations:
	\begin{itemize}
		\item Method \texttt{PUT} \emph{creates} a new resource from a transferred representation.
		\item The idempotent method \texttt{GET} \emph{reads} a representation of a resource's current state.
		\item Method \texttt{POST} \emph{updates} the state of a resource to the transferred representation.
		\item Method \texttt{DELETE} \emph{deletes} a resource.
	\end{itemize}

	\item \emph{Self-descriptive messages.} \ \ A resource can be represented in various formats, e.g., HTML, XML, or JSON, so a client can choose or negotiate a viable representation.
	On the other hand, a resource's metadata allow decision making for caching, content negotiation, checksums, authentication, and access control \cite{Pautasso2008}.

	\item \emph{Hypermedia as the engine of application state.} \ Interaction with a resource is stateless and, in terms of interaction patterns, Send-Receive for a client.
	A RESTful service only manages resource state, and the client is responsible for tracking application state, as shown in Fig. \ref{fig:rest}.
	A pure RESTful service in Fielding's definition has no notion of session as common in Web applications; every Send-Receive interaction has to be self-contained.
	
	A client accesses a RESTful service through a single entry point, i.e., a bookmark, and the service returns hypertext as simultaneous presentation of information and control \cite{Fielding2008}.
	The hypertext outlines the choices a client can take at a specific state; choices are basically hyperlinks that point to continuation URIs.
	Interpretation of a resource then depends on its MIME media type.
	Application state is handled completely on the client side, and REST is an extreme case of dynamically typed interface.
	The hypertext-driven presentation enables loose coupling, maximum freedom in resource namespaces, and allows dynamic discovery of resources.
\end{enumerate}

There has been an extensive debate in the literature on REST versus SOAP/WS-* over the years \cite{Pautasso2009a,Pautasso2010,Potti2011}, including a quantitative comparison by Pautasso et al. \cite{Pautasso2008} with respect to degrees of freedom in both architectures.
While SOAP/WS-* standards precisely define how to implement certain properties of a service, REST is a number of principles that outline characteristics that a service needs to satisfy.
This freedom-of-choice in REST has led to many services that claim to be RESTful but are in fact RPC \cite{Fielding2008}.

\paragraph{Technologies.} \ In comparison with SOAP/WS-* Web services, technologies for REST can also be grouped into four categories of purpose:

\begin{itemize}

	\item \emph{Service description.} \ APIs for RESTful services are often just documented in natural language on a website, e.g., Open APIs for mashups.
There are two attempts in machine-interpretable RESTful service descriptions: WSDL 2.0 \cite{w3c-wsdl} and the Web Application Description Language (WADL) \cite{w3c-wadl}.
The latest version of WSDL supports non-SOAP messages and has more fine-grained control over HTTP bindings to turn them RESTful.

WADL is a description language for HTTP-based Web applications to enable modeling or automatic stub code generation for RESTful service APIs.
WADL is XML based and allows to describe a set of resources, relationships between resources, available methods, and representations for a resource.
Furthermore, WADL supports XSD and Relax NG as schema languages for XML resource representations.

	\item \emph{Service discovery.} \ Resource discovery in REST is dynamic due to the ``Hypermedia as the Engine of Application State'' principle.
	In Fig. \ref{fig:rest}, the \texttt{cart} resource is the entry point and provides hyperlinks to products, so the client discovers its choices and proceeds to the next internal application state.
	
Discovering a service is then uncovering an entry point, i.e., bookmark or hyperlink.
Entry points can be managed in registries accessible as a service, e.g., Google APIs Discovery Service \cite{google-apidiscovery}, announced by syndication services, e.g., AtomPub, or defined in DNS-based service records \cite{RFC6763}.

	\item \emph{Service interaction.} \ Formats of SOAP/WS-* messages are inherently XML-based and specified in a WSDL.
REST does not restrict representations of resources and allows free usage of existing and custom MIME types.
Popular text-based formats in REST are plain-old XML (POX), JSON, and YAML.
As long as metadata refers to a MIME type supported by the client, the format is valid.

REST is bilateral interaction between client and service; multilateral messaging, like WS-Addressing in SOAP-based Web services, is not standardized.
An application of REST with respect to messaging is a Web-accessible interface for a message-based middleware (discussed in Sect. \ref{sub:mom}).

	\item \emph{Service composition.} \ Today, composition of RESTful services primarily takes place in Web mashups, where resources serve as Web components.
There are also attempts toward composition in JOpera~\cite{Pautasso2009b} and workflow orchestration by BPEL for REST~\cite{Pautasso2009}, BPMN for REST~\cite{Pautasso2011}, or the JavaScript-based \emph{S} language~\cite{Bonetta2012}.

\end{itemize}

\paragraph{Hi-REST and Lo-REST.} \ Pautasso et al. \cite{Pautasso2008} distinguish HTTP-based implementations into Hi-REST and Lo-REST:
Hi-REST uses the four HTTP methods for operations, POX for data serialization, and resources have ``nice'' URIs.
But many Web browsers are limited to HTTP methods \texttt{GET} and \texttt{POST} which restricts available REST operations when integrated in Web applications or mashups.
Lo-REST deals with these restrictions and, as a workaround, exploits an HTTP header or a hidden form field to store the actual REST operation that needs be applied.
This of course affects caching infrastructure.

\paragraph{Security and reliability.} \ For secure exchange of representations, REST relies on SSL/TLS in HTTPS.
Only the connection between client and service is secured this way, there is no standardized end-to-end security in a messaging context, like WS-Security.
Furthermore, REST over HTTP has no standardized reliable delivery nor transaction handling compared to SOAP/WS-* Web services.

\paragraph{Constrained RESTful environments.} \ CoAP \cite{RFC7252} is an alternative to HTTP in constrained environments, where computational power and memory are limited, i.e., the Internet of Things~\cite{RFC6690}.
CoAP is designed to easily integrate with HTTP, e.g., by sharing methods, URIs, and media types, and allows Send-Receive interaction using separate or piggybacked acknowledgments depending on the time needed to prepare a resource representation.
The protocol furthermore supports multicast for One-to-Many Send and One-to-Many Send-Receive messaging and specifies service discovery to implement REST principles in constrained environments.

\paragraph{Software implementations.} \ REST is more an architectural style than a set of technologies, and RESTfulness is achievable in all kinds of Web frameworks or programming environments, e.g., Google Gadgets for mashups \cite{google-gadgets}.

Specifically in the Java world, the JAX-RS API \cite{java-jax-rs} has been introduced for RESTful services, and practically all mentioned SOAP/WS-* software implementations support it too.
Other notable implementations and frameworks are the REST extensions in Microsoft's WCF \cite{ms-wcf}, the PaaS Restlet \cite{restlet}, and the Web Resource Modeling Language (WRML) \cite{wrml} framework for RESTful API design.

\subsubsection{Message-oriented middleware}
\label{sub:mom}

\begin{figure}
	\begin{center}
	\subfigure[Peer-to-peer messaging.]{
\begin{tikzpicture}[<->, >=latex]

\node[draw] (v2) at (-1.7,1.5) {\textsf{Peer}$_1$};

\node[draw] (v5) at (1.2,1.5) {\textsf{Peer}$_4$};
\node[draw] (v7) at (0.5,2.4) {\textsf{Peer}$_3$};
\node[draw] (v6) at (-1,2.4) {\textsf{Peer}$_2$};

\node[draw] (v3) at (0.5,0.6) {\textsf{Peer}$_5$};
\node[draw] (v4) at (-1,0.6) {\textsf{Peer}$_6$};


\draw  (v2) edge (v7);
\draw  (v6) edge (v7);
\draw  (v6) edge (v2);
\draw  (v6) edge (v4);
\draw  (v4) edge (v3);
\draw  (v3) edge (v5);
\draw  (v5) edge (v7);
\draw  (v5) edge (v2);
\draw  (v7) edge (v4);
\draw  (v3) edge (v6);
\draw  (v7) edge (v3);
\draw  (v2) edge (v3);
\draw  (v5) edge (v4);
\draw  (v5) edge (v6);
\draw  (v2) edge (v4);
\end{tikzpicture}
		\label{fig:mom_brokerless}
	}
	\hspace{0.2em}
	\subfigure[Broker-based messaging.]{
\begin{tikzpicture}[<->, >=latex]
\node[draw, align=center] (v1) at (-0.1,0) {\textsf{Message-Oriented}\\ \textsf{Middleware}};
\node[draw] (v2) at (-1.4,0.9) {\textsf{Peer}$_1$};

\node[draw] (v5) at (-0.1,-1) {\textsf{Peer}$_5$};
\node[draw] (v7) at (1.2,0.9) {\textsf{Peer}$_3$};
\node[draw] (v6) at (-0.1,1) {\textsf{Peer}$_2$};

\node[draw] (v3) at (1.2,-0.9) {\textsf{Peer}$_4$};
\node[draw] (v4) at (-1.4,-0.9) {\textsf{Peer}$_6$};

\draw  (v2) edge (v1);
\draw  (v6) edge (v1);
\draw  (v7) edge (v1);
\draw  (v3) edge (v1);
\draw  (v5) edge (v1);
\draw  (v4) edge (v1);
\end{tikzpicture}
		\label{fig:mom_mom}
		\hspace{1em}
	}
	\end{center}
\vspace{-1\baselineskip}
\caption{A message-oriented middleware abstracts communication between heterogeneous peers. A broker can reduce the communication complexity \cite{Curry2005}, and a peer can participate as service, client, or both.}
\label{fig:mom}
\end{figure}
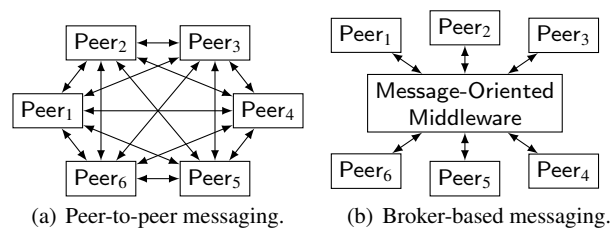

To cope with increasing demands on scalability, flexibility, and reliability, a message-oriented middleware (MOM) is an infrastructure for loosely coupled interprocess communication in an enterprise service bus or clouds \cite{Curry2005}.
Especially in clouds, loose coupling allows to rapidly scale message producers and consumers.
A message with respect to MOM is an autonomous, self-contained entity that models an event and separates into a header and a body or payload.
The middleware provides technical means of exchange, so a peer can exchange messages with other connected peers.

A central concept in MOM is the notion of a \emph{message queue} (or channel) for storing, transforming, and forwarding messages.
Message queues enable asynchronous interaction, and a simple form is a First-In-First-Out (FIFO) queue.
There are two different approaches to MOM using message queues as shown in Fig. \ref{fig:mom}:

\begin{itemize}
	\item \emph{Peer-to-peer messaging.} \ A unified middleware component in every peer coordinates discovery and interaction between peers.
	\item \emph{Broker-based messaging.} \ The middleware acts as a broker to provide a messaging infrastructure between the heterogeneous peers.
\end{itemize}

Peers can participate as client, service, or both \cite{Curry2005}.
A broker reduces the communication complexity between a number of peers but can incur delays in real-time applications because an additional store-and-forward procedure is necessary.

In terms of interaction patterns, a trivial message queue allows bilateral Send and Receive, for asynchronous messaging, and multilateral One-to-Many Send, e.g., publish-subscribe.
Using message queues in a broker architecture allows to implement sophisticated routing patterns.
In general, a MOM is characterized by Curry \cite{Curry2005}:

\begin{itemize}
	\item \emph{Messaging specification.} \ A MOM needs to specify the format of messages and transport mechanisms.
	Interconnecting proprietary MOM systems is achieved through adapters or bridges.	
	
	\item \emph{Message filtering.} \ A core functionality of a MOM is filtering for message delivery.
	Curry \cite{Curry2005} distinguishes:
	\begin{itemize}
		\item A channel-based system offers predefined groups of events as channels, where clients can subscribe to.
		\item Messages in a subject-based system carry metadata in the message header, e.g., a subject. A client subscribes messages, where the metadata matches some given pattern.
		\item In a content-based system, a client subscribes messages, where the message body satisfies a set of properties expressed in a query language. 
		\item Composite events functionality extends a content-based filtering with property matching across sets or sequences of messages.
	\end{itemize}

	\item \emph{Message transformation.} \ Messages can originate from various heterogeneous sources and consequently carry all kinds of content types as payload.
	A MOM can offer APIs to modify messages, e.g., XML transformations.
	
	\item \emph{Integrity, reliability, and availability.} \ A MOM can have properties to increase the overall Quality-of-Service:
	\begin{itemize}
		\item Transactions and Atomic Multicast Notification;
		\item Reliable message delivery: at-least-once, exactly-on\-ce, or at-most-once;
		\item Guaranteed message delivery by acknowledgments;
		\item Prioritization of messages;
		\item Load balancing over several brokers or queues; and
		\item Message broker clustering for fault tolerance.
	\end{itemize}
\end{itemize}

A MOM is typically accessed through an API to abstract the technical details of message exchange.
Due to the transport-agnostic design of SOAP/WS-* services, a MOM can also serve as a transport mechanism for SOAP messages.

\paragraph{Java Message Service.} \ The general purpose API named Java Message Service (JMS)~\cite{oracle-jms} is maintained in a Java community process for MOM support.
JMS defines a number of operations for creating, sending, receiving, and reading messages.
It is transport-agnostic to abstract messaging from MOM implementations and therefore relaxes vendor lock-in.
JMS is a universal interface for interacting with heterogeneous messaging systems \cite{Curry2005}.
A message body is dynamically typed according to the content type information stored in the  header.

Some examples for JMS-enabled software implementations are the JMS reference implementation OpenMQ \cite{openmq}, IBM Websphere MQ \cite{ibm-webspheremq}, or TIBCO Enterprise Message Service \cite{tibco-ems}.

\paragraph{RESTful Messaging Service.} \ The motivation for RESTful Messaging Service (RestMS) \cite{restms} is Web-compatible messaging by using HTTP as transport mechanism and REST principles to describe locations, i.e., URLs, where messages can be posted to and received from.
RestMS is an API specification, where XML-based messages are sent and received using HTTP methods.
With respect to the REST service, resource locations are distinguished into feeds for incoming and pipes for outgoing messages.
Feeds are joined with pipes on the service-side for message distribution.
Message types in RestMS refer to XML, JSON, and a set of MIME content types for dynamically typing data.
The specification also includes profiles to connect to other messaging infrastructures, e.g., AMQP.

\paragraph{Open Middleware Agnostic Messaging API.} \ Due to the diversity in middleware standards and wire formats, the Open Middleware Agnostic Messaging API (OpenMAMA) \cite{openmama} initiative is an attempt to provide a single API for developing applications spanning across multiple MOMs.
For correct translation messages and operations, a MOM has to provide a so-called OpenMAMA bridge implementation.

OpenMAMA is available as open-source library.
It offers a built-in bridge for AMQP-enabled Apache Qpid and supports several bridges for proprietary messaging infrastructures in the finance sector.

\paragraph{Proprietary messaging solutions.} \ MSMQ \cite{ms-msmq} is a MOM for standalone integration or as a transport mechanism in Microsoft's WCF, next to Web services and COM+.
It offers guaranteed message delivery, message routing, transactions, prioritization, and a simple type system for message body types.
When used as a transport in WCF, a message body is either XML, binary, or ActiveX format.
Beside its proprietary protocols, messages can also be transmitted over COM+.
In terms of security, MSMQ allows authentication and encryption of messages.
There is no broker in MSMQ; similar to Fig. \ref{fig:mom_brokerless}, a queue is hosted locally on a peer, and processes can store and retrieve messages.
In terms of service interaction patterns, MSMQ is bilateral Send and Receive.
MSMQ can exploit IP multicast to replicate a message for addressing multiple queues.
A Microsoft alternative with brokerage support is SQL Server Service Broker~\cite{ms-servicebroker}.

Other proprietary MOM software products are the brokerless TIBCO Rendezvous \cite{tibco-rendezvous}, which uses direct connections between peers similar to MSMQ, Oracle Tuxedo Message Queue \cite{oracle-tuxedomq} as part of the Oracle Tuxedo application server for cloud middleware, and Terracotta Universal Messaging \cite{sag-messaging}.

\paragraph{Advanced Message Queuing Protocol.} \ Historically, MOM solutions have relied on proprietary protocols, and JMS is an attempt to agree on a compatible interface.
Interoperability between varying MOM solutions is still difficult; costly JMS adapters or bridges are necessary to connect different transport mechanisms.
AMQP \cite{oasis-amqp} unifies messaging through an agreed-on wire format and has a similar role like HTTP in Web applications.
While the OASIS AMQP 1.0 standard is restricted to the transport model for interoperability over the Internet, messaging architectures are specified by the AMQP working group \cite{oasis-amqp-old}.

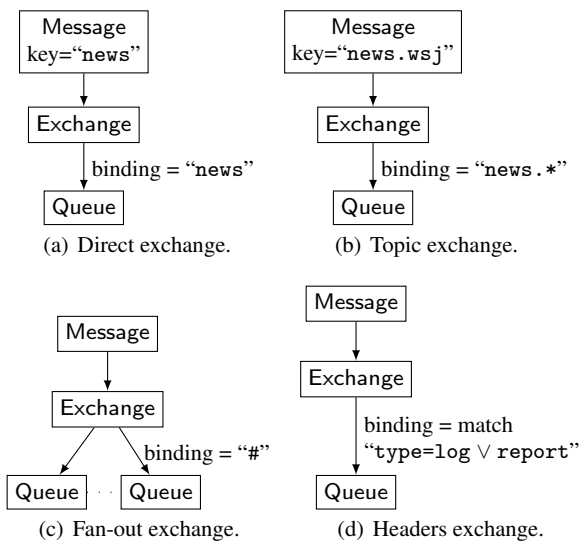
\begin{figure}
	\begin{center}
	\subfigure[Direct exchange.]{
\usetikzlibrary{decorations.pathreplacing}
\begin{tikzpicture}[->, >=latex]
\tikzstyle{end}=[draw]
\node[end] (v1) at (0,-1.1) {\textsf{Queue}};
\node[end] (v2) at (0,0) {\textsf{Exchange}};
\draw  (v2) edge node[right,yshift=-0.5mm] {binding = ``\texttt{news}''} (v1);
\node[end, align=center] (v3) at (0,1.1) {\textsf{Message}\\key=``\texttt{news}''};
\draw  (v3) edge (v2);
\end{tikzpicture}
		\label{fig:amqp_direct}
	}
	\subfigure[Topic exchange.]{
\usetikzlibrary{decorations.pathreplacing}
\begin{tikzpicture}[->, >=latex]
\tikzstyle{end}=[draw]
\node[end] (v1) at (0,-1.1) {\textsf{Queue}};
\node[end] (v2) at (0,0) {\textsf{Exchange}};
\draw  (v2) edge node[right,yshift=-0.5mm] {binding = ``\texttt{news.*}''} (v1);
\node[end, align=center] (v3) at (0,1.1) {\textsf{Message}\\key=``\texttt{news.wsj}''};
\draw  (v3) edge (v2);
\end{tikzpicture}
		\label{fig:amqp_topic}
	}
	\subfigure[Fan-out exchange.]{
\usetikzlibrary{decorations.pathreplacing}
\begin{tikzpicture}[->, >=latex]
\tikzstyle{end}=[draw]
\node[end] (v1) at (0.7,-1.1) {\textsf{Queue}};
\node[end] (v2) at (0,0) {\textsf{Exchange}};
\draw  (v2) edge node[right,yshift=-0.5mm,xshift=0.2mm] {binding = ``\texttt{\#}''} (v1);
\node[end, align=center] (v3) at (0,1) {\textsf{Message}};
\draw  (v3) edge (v2);
\node[draw] (v4) at (-0.8,-1.1) {\textsf{Queue}};
\draw  (v2) edge (v4);
\draw[-, loosely dotted]  (v4) edge (v1);
\end{tikzpicture}
		\label{fig:amqp_fanout}
	}
	\subfigure[Headers exchange.]{
\usetikzlibrary{decorations.pathreplacing}
\begin{tikzpicture}[->, >=latex]
\tikzstyle{end}=[draw]
\node[end] (v1) at (0,-1.5) {\textsf{Queue}};
\node[end] (v2) at (0,0) {\textsf{Exchange}};
\draw  (v2) edge node[right,yshift=-0.5mm, align=left] {binding = match\\``\texttt{type=log} $\lor$ \texttt{report}''}(v1);
\node[end, align=center] (v3) at (0,1) {\textsf{Message}};
\draw  (v3) edge (v2);
\end{tikzpicture}
		\label{fig:amqp_headers}
	}
	\end{center}
\vspace{-1\baselineskip}
\caption{AMQP defines four types of exchanges. A producer creates a message and sends it to an exchange. Depending on the exchange type and bindings, the message is delivered to queues, where consumers can fetch it from.}
\label{fig:amqp}
\end{figure}

The AMQP specification distinguishes a transport model (discussed in Sect. \ref{sub:MessageingProtocol}) and a queuing model \cite{OHara2007}.
The semantic queuing model defines terms like message, queue, exchange, and binding with respect to AMQP.
Messages always end up in queues which are analogous to postal mailboxes.
A queue stores messages and offers functionality for searching, reordering, or transaction participation.
If a client wants to send a message, it chooses a broker-like exchange which is responsible for delivering messages to queues.
An exchange can be offered as a service, and there exists an individual URI scheme (\texttt{amqp:} or \texttt{amqps:}) \cite{rabbitmq-amqpuri} to locate an exchange.
A binding is a set of queue-specific arguments for an exchange.
As shown in Fig. \ref{fig:amqp}, there are different exchange types with respect to message filtering capabilities~\cite{OHara2007}:

\begin{itemize}
	\item In a direct exchange, a message has a routing key and is sent to the queue, whose binding is equivalent to the routing key.
	In case of multiple queues with identical bindings, multiple message copies are delivered, i.e., a channel-based system.
	\item A topic exchange forwards copies of a message to all client queues, where the message routing key matches a queue's binding pattern, i.e., a subject-based system for publish-subscribe delivery.
	\item In a fan-out exchange, messages are forwarded to a set of queues without a specified binding, i.e., channel-based system.
	\item A headers exchange matches the headers of a message against predicate arguments of client queues beyond the routing key, i.e., a content-based system.
\end{itemize}

Messages are finally fetched from queues by consumer processes.
AMQP provides guaranteed delivery, authentication, wire-level encryption, and transaction-based messaging for reliability.
In terms of patterns, an exchange applies pattern Send in case of direct delivery or One-to-Many Send in other cases.
Due to the self-contained type system and self-describing message content, messages are dynamically typed in AMQP.

Examples for JMS-compatible broker implementations are OpenAMQ \cite{imatix-openamq}, JORAM \cite{joram}, WSO2 Message Broker \cite{wso2-broker}, SwiftMQ \cite{swiftmq}, Apache Qpid \cite{apache-qpid}, and Red Hat Enterprise MRG \cite{redhat-mrg}.

\paragraph{Extensible Messaging and Presence Protocol.} \ While the XMPP \cite{RFC6122,RFC6120,RFC6121} has been intended as an open standard for instant messaging, presence information, and contact list maintenance in chat applications, it also has middleware properties.
In its base specification, XMPP exchanges messages as XML stanzas in client-to-service and service-to-service communication for federated services.
An XMPP service therefore takes the role of a broker.

XMPP is particular attractive for MOM scenarios, where Web agents are involved because it supports HTTP as transport mechanism and most Web browsers and JavaScript runtime environments are capable of processing XML stanzas.
Furthermore, XMPP is also considered as a suitable messaging protocol for Internet of Things applications \cite{xmpp-iot}.
The protocol is extensible and extensions are specified in a community process.
MOM-specific extensions are:

\begin{itemize}
	\item Transfer of Base64-encoded binary content with an assigned MIME media type \cite{xep-binary};
	\item RPC over XMPP \cite{xep-rpc};
	\item Service discovery \cite{xep-sd};
	\item Publish-subscribe \cite{xep-pubsub} for broker scenarios, extended addressing \cite{xep-addressing} for message routing, and event notification extensions \cite{xep-event,xep-eventpersonal};
	\item Reliable message transport \cite{xep-msgprocessing}; and
	\item SSL/TLS protected transport mechanism and S/MIME \cite{RFC5751} for end-to-end message encryption.
\end{itemize}

By default, messages in XMPP are XML stanzas and bodies are restricted to text only; there exists a notion of message type, but it is limited to instant messaging applications.
Therefore, out-of-band signaling or a custom protocol, e.g., XMPP bits of binary \cite{xep-binary}, is required to discover message content types in a middleware scenario.

XMPP can also serve as a messaging infrastructure for SOAP/WS-* Web services \cite{xep-soap}.
Beside instant messaging, XMPP has been successfully deployed in the VIRTUS middleware for Internet of Things applications \cite{Conzon2012} using the real-time collaboration server software OpenFire \cite{openfire}.
Another software that offers XMPP messaging over WebSocket is the Kaazing WebSocket Gateway \cite{kaazing}.

\paragraph{Streaming Text Oriented Messaging Protocol.} \ The simple text-based wire protocol STOMP \cite{stomp} is for asynchronous message exchange between a client and a service or broker with simplicity and interoperability in mind.
In the open standard of STOMP, a client and a service establish a session and asynchronously exchange frames of type \texttt{Message}, \texttt{Receipt}, or \texttt{Error}; a frame is partitioned into a command, header fields for metadata, and content of a certain MIME type.
Messages are therefore dynamically typed.
The protocol supports transactions and acknowledgments for reliable message delivery.

STOMP supports either bilateral messaging, i.e., Send and Receive, or broker-based publish-subscribe as in Fig. \ref{fig:pubsub_broker} for One-to-Many Send interaction.
Two notable service implementations are CoilMQ \cite{coilmq} and, for the latest protocol version 1.2, Stampy \cite{stampy}.

\paragraph{Message Queue Telemetry Transport.} \ MQTT \cite{ibm-mqtt} originates from IBM and is now an open OASIS standard \cite{oasis-mqtt} for lightweight machine-to-machine messaging and Internet of Things applications, where bandwidth is limited.
MQTT is intended for broker-based publish-subscribe architectures, as shown in Fig. \ref{fig:pubsub_broker}, i.e., One-to-Many Send interaction.
An MQTT message can encapsulate binary payload up to 256 megabytes, but there is no notion of content type.
The participating parties therefore have to agree on allowed formats out-of-band.
For reliability, the protocol offers acknowledgments and retransmissions, but there is no transaction functionality.

Two notable MQTT broker software implementations are HiveMQ \cite{hivemq} and Mosquitto \cite{mosquitto}.
Both support Web clients using WebSocket.
Another application that relies on MQTT messaging is Facebook Messenger \cite{Zhang2011}.

\paragraph{Data Distribution Service for real-time systems.} \ The open standard DDS~\cite{omg-dds} specifies a machine-to-machine MOM for publish-subscribe message distribution, real-time message delivery, scalability, and high throughput.
Fields of application include the finance and defense sector, industry, aerospace, Internet of Things, and mobile devices \cite{twinoaks-dds}.

Contrary to MQTT, DDS facilitates a data-centric, peer-to-peer interaction in the spirit of Fig. \ref{fig:mom_brokerless}.
A \emph{domain} partitions entities such as publisher, subscriber, and topic.
A topic in a domain has a unique name and a strong datatype for publishing; these types are specified in an IDL, and messages are therefore statically typed.
Subscribers in the domain request data via the topic, and publishers in the domain are responsible for message distribution \cite{oci-devguide}.

DDS supports rich Quality-of-Service policies for data transmission.
Interoperability between software implementations is achieved by the RTPS \cite{omg-ddsi} wire protocol.
To locate endpoints of peers, DDS provides dynamic discovery of publishers, subscribers, topics, and datatypes with respect to topics \cite{twinoaks-dds}.
Reliable message delivery is achieved by negative acknowledgment when data is missing \cite{oci-devguide}.
Security extensions for DDS, e.g., encrypted transport, are still in a beta state at time of writing \cite{omg-ddssec}.

Notable software implementations are OpenDDS~\cite{oci-opendds}, RTI Connext DDS~\cite{rti-dss}, PrismTech OpenSlice DDS~\cite{prismtech-dds}, and Twin Oaks CoreDX DDS~\cite{twinoaks-coredx}.

\paragraph{Apache Kafka.} \ Developed by LinkedIn, Apache Kafka \cite{apache-kafka} is a message broker specification and implementation for high-throughput publish-subscribe messaging, i.e., One-to-Many Send interaction.
Kafka has an individual binary wire format protocol on top of TCP, and for fault tolerance, it supports clustering of brokers, persistent storage, and replication of messages.

On a conceptual level, Kafka distinguishes between topics for messages, producers that publish messages, and consumers that subscribe to topics.
For every topic, a Kafka cluster maintains a partitioned log, where every partition stores an ordered sequence of published messages.
The messages are kept for a configurable timespan, and partitions are replicated and distributed over servers in the Kafka cluster for fault tolerance and performance.
The distributed log in Kafka guarantees the ordering of published and consumed messages in a certain topic. 
For a subscribed topic, a consumer maintains an offset in the message sequence to keep track of already processed ones.
Through this offset, a consumer can also access older messages if they are still available on the cluster.

A message body is a byte sequence of a certain length and has no notion of type.
Content type information therefore needs to be agreed out-of-band or by using a custom protocol.
An interface for Web clients to subscribe to Kafka over WebSockets is already in an experimental state \cite{Black2014}.

\paragraph{ZeroMQ.} \ The intelligent socket library ZeroMQ \cite{imatix-zeromq} aims for more flexible connectivity between peers.
ZeroMQ offers several network transports, including TCP, UDP, and IP multicast, and a number of sockets types for architectural patterns.
Messages are delivered to a thread- or process-local queue and made available through a socket.
The specification defines the following socket types:

\begin{itemize}
	\item \texttt{REQ} and \texttt{REP} for bilateral Send-Receive;
	\item \texttt{DEALER} and \texttt{ROUTER} for routing patterns;
	\item \texttt{PUB} and \texttt{SUB} for publish-subscribe One-to-Many Send;
	\item \texttt{PUSH} and \texttt{PULL} for workload distribution through One-to-Many Send and One-from-Many Receive;
	\item \texttt{PAIR} for asynchronous Send or Receive between two sockets.
\end{itemize}

ZeroMQ has no notion of broker because it is a socket abstraction.
However, a MOM broker could be implemented using ZeroMQ.
Messages are sequences of bytes and do not have a specified content type.
The content type needs to be agreed on out-of-band or requires a custom protocol.

An attempt to provide ZeroMQ access in Web environments is NullMQ \cite{nullmq}.
The JavaScript library uses WebSockets and a modified version of STOMP to bridge ZeroMQ messages into Web browsers.

ZeroRPC \cite{dotcloud-zerorpc} integrates RPC on top of ZeroMQ.
Information is serialized as JSON-based MessagePack format and forwarded over ZeroMQ connections.
A service interface is dynamically typed, and an ZeroRPC has been used in the dotCloud PaaS.

\paragraph{Polyglot message brokers.} \ A natural approach for interconnecting several MOM standards is polyglot message brokerage.
Three notable JMS-compliant software implementations in this area are Apache ActiveMQ \cite{apache-activemq}, RabbitMQ \cite{rabbitmq}, and JBoss HornetQ \cite{hornetq}.

Beside features for scaling and clustering, the messaging core of Apache ActiveMQ, referred to as Apollo \cite{apache-activemq-apollo}, uses the OpenWire~\cite{apache-openwire} wire format, but also supports standards like AMQP, MQTT, and STOMP over WebSockets.
ActiveMQ furthermore provides a proprietary HTTP-based RESTful API for Web clients.

RabbitMQ supports AMQP, STOMP, MQTT, and also HTTP as transport.
Messages over HTTP can be sent in three ways: a native Web management API, STOMP over WebSockets, and JSON-RPC for Web browser integration.

HornetQ \cite{hornetq} is a MOM that originates from the JBoss application server. 
It supports AMQP, has an HTTP-based RESTful Web interface, and provides STOMP over WebSockets for Web clients.

\paragraph{Message queuing as a service.} \ Message brokerage has become an attractive cloud service.
A broker is a critical component in a MOM architecture and needs fault-tolerance, regular maintenance, and scalability; a message queue cloud service can eventually reduce cost.
Amazon Web Services offers Simple Queue Services (SQS) \cite{amazon-simplequeue} for transporting untyped text-based messages up to 256 kilobytes.
SQS operates on a SOAP/WS-* Web service stack accessible through HTTP and HTTPS bindings.

Google's App Engine offers Pull Queues \cite{google-pullqueues} and Push Queues \cite{google-pushqueues} for messaging and App Engine task distribution.
Both queue types are accessible through a RESTful API and use JSON format for messages.
While Pull Queues need to be polled, Push Queues rely on webhooks for HTTP-based message delivery.
Google has also announced Cloud Pub/Sub \cite{google-pubsub}, a broker-based publish-subscribe messaging service for the App Engine, cloud apps, and Web clients.
Using a RESTful API, Cloud Pub/Sub distributes JSON-based messages according to topics.
Subscribers can either poll for new messages or register a webhook for notification.
The service supports guaranteed message delivery by maintaining a queue for every subscriber, and messages are removed from the queue, when the client acknowledges the message.

Microsoft also offers two cloud-based messaging solutions: Azure Queues and Service Bus Queues \cite{ms-azurequeue}.
Azure Queues provide direct messaging between cloud services, and they are accessible through a RESTful interface.
Messages are sequences of bytes and therefore not typed similar to Amazon SQS.
Service Bus Queues offer advanced architectures such as publish-subscribe and routing patterns.
Windows applications and peers can access a service bus through WCF or directly by HTTP.
A \texttt{BrokeredMessage} in a Service Bus Queue explicitly refers to a user-specified message body content.
Service Bus Queues also offer an AMQP interface \cite{ms-amqp}.

Two cloud services that offer AMQP brokerage as a service are StormMQ \cite{stormmq} and IronMQ \cite{ironmq}.
CloudAMQP specifically offers the polyglot broker RabbitMQ as a Service \cite{cloudamqp}.
CloudMQTT \cite{cloudmqtt} is another pay-per-use broker for MQTT messaging, e.g., for complex event processing in Internet of Things environments.
Rackspace Cloud Queues \cite{rackspace-queues} supports publish-subscribe architectures by a HTTP-based RESTful API in the spirit of RestMS.

\section{Discussion}
\label{sec:discussion}

\begin{table}
\caption{Summary of interaction patterns in communication protocols.}
\label{tab:protocols}
\scriptsize
\begin{tabular}{p{0.21\columnwidth}|p{0.69\columnwidth}}
\hline\noalign{\smallskip}
Protocol & Interaction patterns and properties \\
\noalign{\smallskip}\hline\hline\noalign{\smallskip}
IPv4, IPv6 & Bilateral Send and Receive, multilateral One-to-Many Send (IP multicast and broadcast) \\
TCP & Bilateral Send and Receive, bidirectional byte streams with delivery and order guarantees \\
UDP & Bilateral Send and Receive, multilateral One-to-Many Send (IP multicast) \\
MPTCP & Bilateral Send and Receive, multihoming and multiplexing over individual TCP connections \\
SCTP & Bilateral Send and Receive, multihoming, multiplexing, byte-oriented messages, optional delivery and order guarantees \\
QUIC & Bilateral Send and Receive on top of UDP, multiplexing \\
\noalign{\smallskip}\hline\noalign{\smallskip}
HTTP, HTTPS & Client-to-service Send-Receive \\
Comet & Multi-Responses from service to client, exploits HTTP persistent connections \\
Reverse HTTP & Client-side Receive-Send, switched roles after an HTTP Send-Receive cycle\\
BOSH & Bilateral Send and Receive, two TCP connections for HTTP, hanging request for service Send \\
SSE & Multi-Responses from service to client, similar to Comet \\
WebSocket & Bilateral Send and Receive, similar to TCP, established in an HTTP Send-Receive cycle \\
SPDY & Client-to-service Send-Receive, optimization of HTTP wire format, Multi-Responses \\
WebRTC & Client-to-client Send and Receive, SCTP with DTLS \\
CoAP &  UDP-based alternative to HTTP in the Internet of Things, client-to-service Send-Receive, One-to-Many Send, One-to-Many Send-Receive\\
\noalign{\smallskip}\hline
\end{tabular}
\end{table}

Tables~\ref{tab:protocols},~\ref{tab:architectures1} and \ref{tab:architectures2} summarize interaction patterns and properties of discussed communication protocols and architectures.
These summaries provide an overview of the state-of-the-art for practitioners, support them in modeling of a cloud architecture guided by interaction requirements, and therefore bridge the gap between conceptual patterns and today's protocols and architectures.

Communication protocols are primarily designed for bilateral interaction between two peers.
Protocols for multilateral interaction, i.e., UDP over IP multicast, have applications in media streaming and grid computations, but multicast is typically disabled by today's cloud providers and workarounds are needed \cite{Mitchell2014}.

Multilateral interaction between peers is achieved on an architectural level.
A prominent architecture in the survey is broker-based One-to-Many Send interaction, i.e., publish-subscribe.
Routing patterns typically arise in low-level IP packet routing and high-level messaging, e.g., SOAP with WS-Addressing, and content-based MOM filtering.

\begin{table}
\caption{Summary of interaction patterns in Web architectures.}
\label{tab:architectures1}
\scriptsize
\begin{tabular}{p{0.21\columnwidth}|p{0.69\columnwidth}}
\hline\noalign{\smallskip}
Protocol & Interaction patterns and properties \\
\noalign{\smallskip}\hline\hline\noalign{\smallskip}
Web application & HTTP or HTTPS Send-Receive interaction, Multi-Responses for audio and video streaming, push technology and AJAX for asynchronous Send and Receive, dynamically typed \\
Webhook & Service-to-service Send between Web applications, callback URLs, for Web routing patterns \\
Web feed & One-to-Many Send, i.e., publish-subscribe, service-to-client: by client polling or push technology, service-to-service: by webhooks \\
Mashup & Multiple simultaneous Web interactions \\
\noalign{\smallskip}\hline
\end{tabular}
\end{table}

\subsection{Cloud computing aspects}
 
Economies of scale, pay as you go, and service delivery over the Internet are three major aspects of cloud computing \cite{Foster2008}.
Infrastructure-as-a-Service is not part of this study; however, Bitar et al. \cite{Bitar2013} review the state-of-the-art technologies in provider-side cloud networking.
The state-of-the-art of protocols and architectures for PaaS and SaaS is presented in the previous sections.

While message passing has turned enterprise services scalable, the custom messaging wire formats are not as reliably forwarded over the Internet as Web protocols.
Beside HTML5-enabled Web 2.0, RPC, SOAP/WS-*, and REST for cloud services, there is a trend towards message passing on top of a Web protocol stack, e.g., RESTful APIs for message brokers, XMPP over HTTP or WebSocket, SOAP over XMPP, STOMP over WebSocket, MQTT over WebSocket, Kafka over WebSocket, and NullMQ to name a few.
In fact, AMQP over WebSocket \cite{oasis-amqpws} is an ongoing standardization effort by OASIS, so Web clients can be directly connected to a message bus.
XMPP already has social aspects from its instant messaging origin and is therefore a good candidate for cloud messaging applications~\cite{Flores2013,Hornsby2010}.
This trend towards Web-compatible messaging differentiates cloud service technology from traditional MOM.

With respect to fault tolerance, Wenbing et al.~\cite{Wenbing2010} introduce a Low Latency Fault Tolerance (LLFT) middleware that specifies a messaging protocol based on UDP over IP multicast.
Services are replicated in virtual groups, a leader is selected, and the LLFT protocol assures reliable, ordered group-to-group message delivery.

Inter-cloud communication and cloud federation also require protocols and architectures to scale, share, and synchronize resources or to authenticate and authorize peers.
Two state-of-the-art reviews of inter-cloud computing are given by Toosi et al.~\cite{Toosi2014} and Grozev and Buyya~\cite{Grozev2014}.
Examples for inter-cloud protocols include XMPP messaging~\cite{Bernstein2009,Celesti2010} (which is also a foundation for standardized inter-cloud interoperability \cite{ieee-p2302}), AMQP \cite{Bernstein2009}, and SOAP/WS-* Web services \cite{Buyya2010}.
Furthermore, Lloret et al. \cite{Lloret2014} propose a custom TCP-based protocol for fault tolerance and load distribution in a distributed inter-cloud scenario.

\paragraph{Cloud to device messaging.} \ A cloud push service is conceptually a message broker that forwards notification messages from third-party services to apps on mobile devices, e.g., smartphones.
The notification client center on the mobile device registers at the cloud broker and typically relies on HTTP push technology to receive near-real-time notifications, often in JSON format.

Today's prominent standards are: Google Cloud Messaging (GCM)~\cite{google-gcm}, Apple Push Notification (APN)~\cite{apple-push}, Windows Push Notification Services (WNS)~\cite{ms-push}, Blackberry Push Service~\cite{blackberry-push}, and
Nokia Notifications~\cite{nokia-notifications}.

\begin{table}
\caption{Summary of interaction patterns in service architectures.}
\label{tab:architectures2}
\scriptsize
\begin{tabular}{p{0.21\columnwidth}|p{0.69\columnwidth}}
\hline\noalign{\smallskip}
Protocol & Interaction patterns and properties \\
\noalign{\smallskip}\hline\hline\noalign{\smallskip}
XML-RPC & HTTP Send-Receive, dynamically typed \\
JSON-RPC & Send-Receive, dynamically typed \\
Apache Avro & Send-Receive, dynamically typed\\
Apache Etch & TCP Send-Receive, dynamically typed\\
Apache Thrift & Send-Receive, statically typed\\
Protocol Buffers & Send-Receive, statically typed\\
Hessian, Burlap & HTTP Send-Receive, dynamically typed\\
\noalign{\smallskip}\hline\noalign{\smallskip}
SOAP & Send, Receive, Send-Receive, statically typed\\
WS-Adressing & Multilateral interaction and routing patterns, e.g., Relayed Request\\
WS-Eventing, WS-Notification & One-to-Many Send or One-from-Many Receive, e.g., publish-subscribe\\
\noalign{\smallskip}\hline\noalign{\smallskip}
REST & Send-Receive, i.e., HTTP or CoAP, dynamically typed\\
\noalign{\smallskip}\hline\noalign{\smallskip}
JMS & Transport-agnostic API, dynamically typed\\
OpenMAMA & Middleware-agnostic messaging API\\
RestMS & REST-based messaging API\\
\noalign{\smallskip}\hline\noalign{\smallskip}
JMS-compatible & OpenMQ, IBM Websphere MQ, TIBCO Enterprise Message Service\\
Proprietary & TIBCO Rendezvous, Oracle Tuxedo, Terracotta Universal Messaging \\
MSMQ & Peer-to-peer, Send, Receive, One-to-Many Send (using network multicast), broker-based: SQL Server Service Broker \\
\noalign{\smallskip}\hline\noalign{\smallskip}
AMQP & Broker-like exchanges, Send, Receive, One-to-Many Send, dynamically typed\\
XMPP & Broker-based Send and Receive; with extensions One-to-Many Send, i.e., publish-subscribe, untyped or dynamically typed\\
STOMP & Send and Receive, broker-based One-to-Many Send, i.e., publish-subscribe, dynamically typed\\
MQTT & Broker-based One-to-Many Send, untyped\\
DDS & Peer-to-peer One-to-Many Send, i.e., publish-subscribe, statically typed\\
Apache Kafka & Broker-based One-to-Many Send, i.e., publish-subscribe, untyped\\
ZeroMQ & Untyped socket abstraction, various patterns\\
\noalign{\smallskip}\hline
\end{tabular}
\end{table}

\paragraph{Client-cloud interaction examples.} \ The first cloud example concerns management of resources.
The Open Cloud Computing Interface (OCCI)~\cite{occi-rest} is a standardization attempt toward a unified Web-compatible RESTful API in particular for IaaS.
For example, the OpenStack \cite{openstack} IaaS cloud platform supports OCCI to deploy or scale virtual machines, change network addresses, or allocate storage.

The second example is the Open-Source API and Platform for Multiple Clouds (mOSAIC)~\cite{Petcu2013} that also implements OCCI.
The mOSAIC software platform provides a vendor-neutral way of running so-called cloudlets, i.e., event-driven and stateless components that implement application functionality.
Core components of the mOSAIC platform are responsible for scheduling, monitoring, scaling, and deployment of cloudlets, and the platform provides an asynchronous message bus, e.g., AMQP or Amazon SQS, so cloudlets can interact.
Furthermore, a cloudlet implements connectors and drivers for communication technologies to interact with clients and services.

For example, a simple XML processing service could be implemented in three mOSAIC cloudlets.
The first cloudlet offers a RESTful Web service, where clients can post XML data, and documents are forwarded to the second cloudlet.
When the second cloudlet is notified, it processes the XML according to the business logic and forwards the results to the third cloudlet which implements a database for storage.
The platform can then duplicate cloudlets for scaling during peak loads.

\subsection{Observations}

\paragraph{Multiplexing.} \ There is a trend toward multiplexed transport protocols in the Web to minimize overhead when multiple resources are requested in parallel by a Web client, e.g., SPDY and the upcoming HTTP/2.
SPDY is already available in Google Chrome and effectively used as transport mechanism, when Google cloud services are consumed.
Multiplexing can also contribute to service-oriented architectures in cloud backend infrastructure, e.g., RPC or MOM, when interactions between two peers are highly parallel, e.g., the Mux protocol in Twitter's Finagle \cite{twitter-finagle} RPC framework.

\paragraph{Multihoming.} \ Mobile devices are still a growing market, and an increasing number of devices is simultaneously connected to multiple networks, e.g., Wi-Fi and 3G/4G networks.
Multihoming protocols like MPTCP already exploit this connectivity in Apple devices to increase client-side access quality for the Apple Siri service \cite{apple-mtcp}.
Now that implementations and libraries are available on a popular platform, multihoming will eventually become mainstream.
Experimental Linux kernels with MPTCP support for Android devices do already exist \cite{webrtc-android}.

\paragraph{Pervasive encryption.} \ Middleboxes for various networking applications, e.g., firewalls or traffic shapers, monitor communication all over the Internet, and they often rely on DPI.
Modern transport protocols such as MPTCP and QUIC, but also SPDY proclaim pervasive encryption, so middleboxes cannot tamper with network interaction.
It is safe to assume that network traffic encryption will flourish in the future, and DPI-based monitoring is eventually affected.

Furthermore, an observation in this study is an increasing distrust in X.509 public key infrastructure.
Any certificate authority, trusted by a Web client, can issue a trusted cryptographic certificate for any service FQDN.
Certificate authorities are typically shipped with software, so the trust relationship is hardly reviewed.
A malign authority can issue forged X.509 certificates for man-in-the-middle attacks in network traffic monitoring \cite{Huang2014}.
An observed trend is \emph{certificate pinning} \cite{owasp-pinning}, where a client gets access to additional information to verify an X.509 service certificate, e.g., a fingerprint of the expected public key or a certificate-specific restrictions on authorities.
These techniques are already put to use today, e.g., in mobile phone apps as a protection against reverse engineering \cite{Osborne2012} or as a security mechanism in the Google Chrome browser~\cite{google-pinning}.

\paragraph{Importance of DNS.} \ The majority of architectures in this article rely on DNS as an abstraction layer for locating service endpoints, i.e., host names.
DNS SRV records even allow dynamic service discovery \cite{RFC6763}.
The importance of DNS security aspects is well known, and mainstream adaption of DNSSEC is therefore expected \cite{Herzberg2013}.


\paragraph{Correct types.} \ \ Dynamically typed interfaces offer great flexibility, but rely on correctness of the content type descriptor of a message, so a correct module for parsing is chosen during runtime.
MIME types can be ambiguous because of implicit subtype relations, e.g., every content that satisfies \texttt{application/xml} also satisfies MIME type \texttt{text/plain} but semantics are different.
Nevertheless, Web technology relies on MIME types as identifiers, and incorrect type information can severely affect function and security.

Two problems in Web browsers with respect to type correctness have been observed.
If the \texttt{Content-Type} is missing in an HTTP response, a Web browser tolerates the error and tries to guess the MIME type using heuristics, i.e., content sniffing \cite{Barth2009}.
Furthermore, HTML and XHTML allow to define an expected MIME type as attribute of \texttt{embed} or \texttt{object} markup tags.
There is a type conflict if the expected type and the HTTP response content type are not the same, and the browser is responsible for resolving the conflict.
Type heuristics and conflicting types have already been exploited to attack Web browsers and undermine content policies \cite{Barth2009,Magazinius2013a}.
Correctness of types is therefore essential for secure composition of services.

\paragraph{Polyglot messaging.} \ The large number of messaging standards for MOM complicates access from Web clients.
Due to varying application profiles in MOM, e.g., Internet of Things or enterprise service bus, the diversity of technologies for MOM will likely stay.
Nevertheless, there is a trend toward messaging in Web and cloud clients for near-real-time communication, so an increasing number of software implementations add Web-compatible interfaces like SOAP, JSON-RPC, RestMS, or individual RESTful interfaces.
The survey indicates that implementations tend to offer adapters for Web clients using WebSockets.
So, Web clients can be connected to messaging infrastructure using, e.g., AMQP, STOMP, MQTT, and XMPP.

\paragraph{Client-to-client messaging.} \ WebRTC has the potential to fundamentally change the Web experience.
Today, Web interaction takes place between a Web client and a Web application or service.
A growing amount of user-provided Web 2.0 content needs to be stored and forwarded, and a lot of effort has been invested to deliver media in time, e.g., Content Delivery Networks (CDNs).
Pay-per-use and flexible scalability in cloud computing industrialize this process, but governance of data, trust, and cloud vendor lock-in are still unresolved problems for users. 

JavaScript and HTML5 turn Web browsers into virtual machines of the Web, and WebRTC enables direct, asynchronous interaction between them.
A potential application is a new era of peer-to-peer networking and code mobility, where clients can seamlessly become service providers, and cloud services merely participate for discovery and signaling between clients.
Existing MOM standards could be adapted to provide a messaging infrastructure over WebRTC as transport mechanism.
An example for today's WebRTC usage is PeerCDN \cite{peercdn}, a distributed peer-to-peer cache as an alternative to costly CDN infrastructure.

\paragraph{Monitoring implications.} \ From an interaction aspect, DPI network traffic monitoring \cite{Bendrath2011} is a common security policy decision point, e.g., next-generation firewalls, intrusion detection and prevention systems, data loss prevention, breach detection, censorship, or traffic shaping.
The survey indicates the following implications about DPI systems:

\begin{itemize}
	\item To analyze encrypted network traffic, a DPI system has to break the encryption, which is a hard problem or requires misuse of X.509 public key infrastructure to forge certificates.
	Otherwise, only metadata about an interaction are available for policy decision making.
	\item Assuming that a DPI system is capable of breaking the encryption using a man-in-the-middle-attack by misusing an authority to forge certificates, certificate pinning can still prevent communication, even when the misused certificate authority is trusted by the client.
	\item Furthermore, a DPI system is bound to some physical domain.
	Multihoming communication utilizes all available physical paths, where eventually one path is invisible to the DPI system, e.g., a path over 3G/4G networks.
	So, the monitor only has a limited view of the communication which impairs decision making, even when based on metadata or network traffic characteristics.
\end{itemize}

A conclusion is that future security systems need to be situated directly on the client side or service side at communication endpoints or as a component in the client or service.
Otherwise, visibility of communication is not guaranteed with upcoming communication protocols.

\section{Conclusion}
\label{sec:conclusion}

The study presents the state-of-the-art in technologies for interaction between clients and services in Web and cloud environments.
As clients and services are distributed, communication over networks is necessary for interaction.
The survey distinguishes languages, protocols, and architectures: Languages encapsulate information in a transportable format; protocols specify communication channel languages and rules of engagement for low-level interaction; and architectures integrate languages, protocols, and service interaction patterns into blueprints for service delivery.

Architectures are approached in two different directions.
The Web-oriented view highlights the evolution of the Web and technological advancements, e.g., AJAX, HTML5, syndication, and mashups.
The service-oriented view highlights architectures and technologies for services originating from enterprise environments.
The survey indicates that technologies in both views are more and more converging by turning scalable enterprise architectures Web-compatible.

Languages and protocols that originate from Web applications, e.g., HTTP, are becoming an integral part in service and cloud technology because these protocols are supported on a wide range of platforms and correctly forwarded over the Internet.
On the other hand, architectures for large-scale distributed systems, e.g., RPC or message passing, more and more influence how Web applications are implemented and composed, e.g., mashups, to fully utilize the capabilities of an underlying cloud infrastructure.
Especially new technologies like WebSocket and WebRTC have a potential to mainstream sophisticated MOM architectures in the Web.
Furthermore, upcoming protocols and architectures will have an impact on network monitoring, especially on DPI-based systems.
For the future, alternatives like client- and service-centric monitoring approaches need to be considered.

\begin{acknowledgements}
This research has been supported by the Christian Doppler Society.
The author thanks the anonymous reviewers, Roxana Holom, Tania Nemes, Mircea Boris Vleju, and Philipp Winter for the valuable feedback to improve this work.
\end{acknowledgements}

\bibliographystyle{spmpsci}
\bibliography{literature,rfc}
\end{document}